# Blockchain for Unmanned Underwater Drones: Research Issues, Challenges, Trends and Future Directions

NEELU JYOTI AHUJA, ADARSH KUMAR, MONIKA THAPLIYAL and SARTHIKA DUTT
University of Petroleum and Energy Studies, Dehradun, India

TANESH KUMAR
University of Oulu, Finland

DIEGO AUGUSTO DE JESUS PACHECO
Aarhus University, Denmark

CHARALAMBOS KONSTANTINOU
King Abdullah University of Science and Technology (KAUST), Thuwal, Saudi Arabia

KIM-KWANG RAYMOND CHOO
University of Texas at San Antonio (UTSA), San Antonio, TX, USA

Underwater drones have found a place in oceanography, oceanic research, bathymetric surveys, military, surveillance, monitoring, undersea exploration, mining, commercial diving, photography and several other activities. Drones housed with several sensors and complex propulsion systems help oceanographic scientists and undersea explorers to map the seabed, study waves, view dead zones, analyze fish counts, predict tidal wave behaviors, aid in finding shipwrecks, building windfarms, examine oil platforms located in deep seas and inspect nuclear reactors in the ship vessels. While drones can be explicitly programmed for specific missions, data security and privacy are crucial issues of serious concern. Blockchain has emerged as a key enabling technology, amongst other disruptive technological enablers, to address security, data sharing, storage, process tracking, collaboration and resource management. This study presents a comprehensive review on the utilization of Blockchain in different underwater applications, discussing use cases and detailing benefits. Potential challenges of underwater applications addressed by Blockchain have been detailed. This work identifies knowledge gaps between theoretical research and real-time Blockchain integration in realistic underwater drone applications. The key limitations for effective integration of Blockchain in real-time integration in UUD applications, along with directions for future research have been presented.

Keywords: Blockchain, Data Management, Drone Design, Privacy, Security, Unmanned Underwater Drones.

## 1. INTRODUCTION

The rapid developments and continuous enhancement in drone and robotics technologies have already widened the scope of their efficient applicability in a number of key domains of daily life, e.g., the emergence of UAV for various critical IoT applications such as smart healthcare, industrial automation, transportation, military applications, agriculture, rescue missions, and surveillance [1]. The use of drone technology has become heavily crucial in situations/areas where human intervention is not possible for various reasons, e.g., due to safety or unstable conditions [2]. Recently, UUVs [3] have been one of the emerging research areas in the marine domain aiming to find the optimal, secure and cost-efficient technological solutions for various underwater IoT applications, such as underwater/ocean surveillance and monitoring for ocean species, minerals/resources, examining of underwater oil/gas, cables/internet infrastructure, and water contamination among others [4]. Since drone technology is pivotal element for UUVs communication networks, it is, therefore, can also be referred as UUDs.

----------------------------------------------------------------

Author's address: N. J. Ahuja, Department of Systemics, School of Computer Science, University of Petroleum and Energy Studies, Dehradun, India; email: neelu@ddn.upes.ac.in; M. Thapliyal, Department of Systemics, School of Computer Science, University of Petroleum and Energy Studies, Dehradun, India; email: monikathaps22@gmail.com; S. Dutt, Department of Systemics, School of Computer Science, University of Petroleum and Energy Studies, Dehradun, India; email: sarthikadutt@gmail.com; T. Kumar, Centre for Wireless Communication, University of Oulu, Finland; email: tanesh.kumar@oulu.fi; Diego Augusto de Jesus Pacheco, Department of Business Development and Technology (BTECH), Aarhus University, Herning, Denmark; email: diego@btech.au.dk; A. Kumar, Department of Systemics, School of Computer Science, University of Petroleum and Energy Studies, Dehradun, India; email: adarsh.kumar@ddn.upes.ac.in; C. Konstantinou, Computer, Electrical and Mathematical Sciences and Engineering (CEMSE) Division, King Abdullah University of Science and Technology (KAUST), Thuwal, Saudi Arabia; email: charalambos.konstantinou@kaust.edu.sa; K. R. CHOO, University of Texas at San Antonio (UTSA), San Antonio, TX, USA; email: Raymond.Choo@utsa.edu

The development path towards the UUVs and marine research is further supported/complemented with the advent of 5G and beyond systems and the massive boom in underwater robotics technologies [5],[6]. In addition, the role of various internet, networking and communications technologies such as Edge/Fog computing, Blockchain, Network Softwarization/virtualization, AI and AR/VR have become vital in these developments [7], [8]. UUVs environment may consist of a number of diverse IoT sensors, actuators, and devices along with underwater robots, devices and vehicles deployed at various depths of the water to accomplish the desired goals [9]. Thereafter, the gathered information by UVVs can be sent to the nearest ships, base stations or even to the satellite for further processing/analysis and storage. Some of the well-known UUVs are ROVs and AUVs, and their use in underwater communication depends upon the requirements and demands of the particular underwater applications [10], [11].

Apart from the numerous benefits, there are still several obstacles in the way to successfully deployment the UUVs. Sensors, devices, underwater robotics together with other network elements in the underwater communication are subjected to different security threats due to the number of reasons, i.e., resource constrained nature of devices/sensors, hazardous underwater environment, and unreliable network connectivity [12]. For example, data security and integrity are among the significant threats to underwater monitoring. In addition, the mobility of the resource-limited devices underwater brings difficulty in terms of the authentication of the devices [13]. Moreover, low-quality data transmission can be experienced during underwater communication causing the overall process to be less reliable and require high energy dissipation. This will also make the routing of the information from sender node to the sink node much more complex and overall performance of the network is degraded [14].

In addition to the security and privacy issues, the current underwater communication networks have obstacles in terms of providing secure and optimal resource allocation mechanisms to various needed network elements [15]. Since the several sensor nodes in the underwater wireless networks are limited in resources and due to water current, the nodes and devices are continuously changing their location, it is highly important to enable a dynamic and intelligent resource allocation mechanism [16]. In addition, it is necessary to have secure monitoring and tracking of various available resources in the network to utilize them in a more optimal and efficient way. Along with management of network resources, the underwater network is required to provide secure and trusted solutions for processing and storing massive amounts of the data produced from multiple sources [17].

Considering the above highlighted challenges, Blockchain-based network architectures are seen as prominent solutions for underwater communication [18]. Blockchain as a decentralized and distributed technology enabler provides several vital features to underwater IoT applications such as immutability, trusted data sharing and management, monitoring/tracking of various processes and resources, efficient resource allocation, secure routing mechanism, auditability, and traceability. Blockchain allows secure and trusted data sharing and storage during underwater communication without the intervention of any trusted third party [19]. In some cases, the hostile environment in underwater communication requires autonomous decision making due to unreliable connection with the outside base station/devices. Blockchain can provide the needed dynamic and autonomous decision making using the characteristics provided by the smart contract.

### 1.1 Motivation

More than 70% of the earth's surface is occupied by the water and oceans. The climate challenges on the earth's surface can be predicted through analysing temperature and wind patterns of the sea. Moreover, the ocean comprises different kinds of animals and water species that require monitoring to ensure a safe and protected living environment for them [20]. Therefore, underwater surveillance is crucial for examining the pollution that can be hazardous for the sea environment and the water species [21]. Furthermore, Ocean is also considered as a rich source for natural underwater resources, e.g., mineral resources. In the past years, due to lack of suitable technological and other resources, finding and maintaining these natural resources are among one of the challenging tasks. Moreover, the ocean and marine environment have a variety of useful applications in both commercial and military domains [22].

However, the underwater world was mostly less-explored mainly due to the limitations and unavailability of the required enabling technologies and the advanced equipment. For example, traditional underwater communication and monitoring was performed using satellite/ wired communication cables, causing higher costs, lower data rates and bandwidth and longer propagation delays [11]. With the recent digital transformation due to ICT technologies, the research community has started investigating various applications in the marine/ocean sector, e.g., UUDs. Among other technological enablers, Blockchain technology has emerged as one of the keys enabling technology for the UUDs [19]. It can address several key challenges such as security and privacy, trusted data sharing, storage, process tracking, and resource management [5], [12]. Thus, this paper further studies the utilization of Blockchain in different underwater applications, its benefits, and potential challenges.

### 1.2 Contributions

The main contributions of this paper are described as follows:

- An overview of the background and state-of-the-art concepts related to Blockchain and underwater communication are discussed.
- A study of various challenges in the current UUVs systems and corresponding Blockchain-based solutions is analysed.
- A wide variety of Blockchain use cases for UUVs are presented and patterns of knowledge are discovered.
- Efficient utilization of the Blockchain concepts in the design of UUVs is studied.
- Different possibilities in terms of the integration of various enabling technologies with the Blockchain-enabled UUVs are explored.
- Comparative analysis of various studies conducted over blockchain and useful for UUVs is performed and analyzed to identify the research gaps, challenges and future directions.
- Blockchain and UUV-based designs useful in data collection, analysis, interpretations and implementation are explored and discussed.

### 1.3 Related Surveys and Our Contributions

Blockchain and UUD have been the subject of just a few studies in the literature to yet[29]-[31][49]. Uddin et al. [29] discussed that undersea IoT routing protocol will connect with sensors using a secret key that has been exchanged between the sensors. If all of the sensors are made by the same manufacturer, the undersea IoT can be connected via a secure network (using blockchain technology). This study generalizes the usage of blockchain technology with IoUT environment. However, this work does not addressed the UUDs for IoT creation or integration with blockchain technology. Yong et al. [31] looked at the most prevalent security and privacy vulnerabilities that may occur in the IoTs. Several blockchain-based solutions are being examined as a means of alleviating these problems. An integrated framework for incorporating blockchain technology into an IoT system was presented in this work. Likewise, Zhaofeng et al. [49] discussed the importance of blockchain in handling security issues. Blockchain-based data management system is tamper-resistant, adaptive and reconfigurable. It can be used in a number of settings and is ideal for a wide range of applications, including healthcare and government. Edge computing applications benefit from its high degree of security and endurance. Uddin et al. [30] presented the in-depth survey in blockchain technology for IoT. It also addresses issues such as research gaps, roadblocks, and possible solutions. IoT-based eHealth, smart home and smart vehicle applications are covered. Cloud computing and blockchain technology are also explored. It has been observed in literature that there is no study integrating blockchain with UUDs for underwater surveillance, monitoring and other applications. In order to have a clearer picture of the field, we need to reexamine both the notion of blockchain and UUDs. Blockchain supporting drones research literature has been thoroughly analysed using the taxonomy, use cases and theoretical discussion provided in this work. Also discussed in this work are blockchain-based technologies and frameworks, useful for underwater applications, as well as future prospects. Table 1 shows the comparison of our survey with the existing surveys and other studies.

**Table 1:** Comparative analysis of recent studies in blockchain and UUD with our contribution

| Author | Year | Major Objective | A | B | C | D | E | F | G | H | I | J | Major Observations |
|---|---|---|---|---|---|---|---|---|---|---|---|---|---|
| Valavanis et al. [92] | 1997 | In this investigation, 25 AUVs and 11 AUV control architecture systems are examined. Hierarchical, heterarchical, subsumption, and hybrid AUV control architectures are examined. | ✘ | ✓ | ✘ | ✘ | ✘ | ✘ | ✓ | ✓ | ✘ | C | A proposal is presented to develop a common hardware and software platform for an AUV control architecture, with the purpose of standardising both the hardware and the software platform. In its proposal, IS Projekt proposes a hybrid embedded control architecture based on the QNX real-time operating system and the STD 32SBC. The hybrid embedded control architecture is based on the QNX real-time operating system and the STD 32SBC. |
| Roman et al. [101] | 2010 | The findings of an unique research project including the use of structured light laser profile imaging to construct high-resolution bathymetric maps of underwater archaeological sites are presented in this publication. | ✘ | ✘ | ✘ | ✘ | ✘ | ✘ | ✓ | ✘ | ✘ | **I** | The results of a novel research project involving the use of structured light laser profile imaging to create high resolution bathymetric maps of underwater archaeological sites are presented in this paper. Structured light laser profile imaging was used to create high resolution bathymetric maps of underwater archaeological sites. |
| Zhou et al. [40] | 2015 | An upgraded version of the Basagni et al. designed channel-aware routing protocol (CARP) was created in order to enable location-free and greedy hop-by-hop packet forwarding. | ✘ | ✘ | ✓ | ✘ | ✘ | ✘ | ✓ | ✓ | ✘ | I | Simulations demonstrate that our approach may significantly cut communication costs while also increasing the network's capacity. |
| Kao et al. [9] | 2017 | This work provides an in-depth analysis and evaluation of the Internet of Things, with the most significant contributions of this study falling into three categories (practical underwater applications, | ✘ | ✓ | ✓ | ✘ | ✘ | ✘ | ✓ | ✘ | ✘ | S | For the Internet of Things, this work investigated and analysed channel models, which provide the technological underpinnings that allow developers to create reliable |
| | | differentiating UWSNs and TWSNs, and evaluating channel models). | | | | | | | | | | | communication protocols that are scalable (Internet of Things). In addition, this work explores the contrasts between UWSNs and conventional, and assert that the disparities between the two kinds of networks are the most difficult hurdles to overcome. |
| Yong et al. [31] | 2018 | This study looked at the most prevalent security and privacy vulnerabilities that may occur in the Internet of Things, as well as their causes. Incorporate blockchain technology into the Internet of Things by developing a framework that has the ability to provide high levels of trust for IoT data. | ✓ | ✗ | ✗ | ✓ | ✓ | ✓ | ✗ | ✗ | ✓ | C | This study looked at the most prevalent security and privacy vulnerabilities that may occur in the IoT, as well as their causes. Several blockchain-based solutions are being examined as a means of alleviating these problems. An integrated framework for incorporating blockchain technology into an IoT system was given in this article. |
| Jouhari et al. [5] | 2019 | Acoustic and magneto-inductive communications is investigated in this study, since they are presently the two most widely utilised technologies for underwater networking. In particular, it draws attention to the trade-offs that exist between underwater qualities, wireless communication technology, and the overall quality of communication. | ✗ | ✓ | ✓ | ✗ | ✗ | ✗ | ✓ | ✗ | ✗ | S | Despite the fact that the electromagnetic channel used in terrestrial wireless sensor networks has low route losses and is subject to dynamic channel conditions, it is not suitable for use in underwater communication. This is because acoustic communication delivers excellent channel quality while simultaneously limiting the effect of unfavourable environmental circumstances on the communication channel, it has become more popular. Using acoustic communication for wireless power transfer makes it possible to send electricity in an efficient manner. |
| Uddin et al. [29] | 2019 | This study presents results based on a sensor monitoring architecture that includes many layers of sensors. Additionally, this work has included fog and cloud aspects into our approach, which is a layer-based design comprised mostly of fog and cloud features. Data from the IoUT is protected by the implementation of Blockchain technologies that are suited for this application. | ✓ | ✗ | ✓ | ✗ | ✓ | ✓ | ✗ | ✗ | ✓ | I | To route data to the suitable destination, the proposed system grouping sensors into clusters and selecting a cluster head based on the amount of remaining energy and the level of a node, as specified, in order to route data to the right destination, as specified. If all of the sensors are made by the same manufacturer, the undersea IoT routing protocol will connect with them using a secret key that has been exchanged between the sensors. Blockchain technology, which is being used to store the data securely, allows it to be saved on the cloud. |
| Zhaofeng et al. [49] | 2019 | A blockchain-based data management system (known as BlockTDM) has been developed in order to address the issues associated with edge computing. It is a tamper-resistant data management system that is versatile, adaptive, and reconfigurable. It can be used in a number of settings and is ideal for a wide range of applications, including healthcare and government. Edge computing applications benefit from its high degree of security and endurance, which makes it an excellent choice for this kind of computing. | ✓ | ✗ | ✗ | ✓ | ✓ | ✓ | ✗ | ✗ | ✓ | I | On the frontier of computer technology, this work suggests the use of a blockchain-based data management system. Sophisticated data may be protected via multichannel data segmentation and isolation. The mutual authentication protocol includes the administration and implementation of smart contracts, blocks, and transaction data. |
| Walter et al. [111] | 2020 | To employ optical and infrared footage captured by drones to survey the Geysir geothermal region in Iceland. A major purpose of the research is to better understand the geographic distribution and structural design of geysers, hot springs, and hydrothermal vents in order to better understand them. | ✗ | ✓ | ✓ | ✗ | ✗ | ✗ | ✓ | ✓ | ✗ | S | A comprehensive drone-based photogrammetric and infrared mapping research project was carried out in the Geysir geothermal region in Iceland for the first time. The Structure from Motion approach was used to analyse hundreds of close-range photographs, leading in the construction of a centimeter-scale digital elevation and thermal anomaly map, as well as a centimeter-scale digital elevation and thermal anomaly map. |
| Jahanbakht et al [17] | 2021 | When it comes to data transmission methods for the IoT, this work concentrate on the most current technologies and procedures. This is followed by a discussion of BMD processing and analytics using several types of machine learning techniques. In the end, this work look at the architectural issues associated with the IoTs and then recommend a number of interesting avenues for additional research and development. | ✗ | ✗ | ✓ | ✗ | ✗ | ✗ | ✓ | ✗ | ✗ | S | The issue of underwater communication is explored in this article. Acoustic, electromagnetic, and optical waves have all been used in the development of models for underwater communication. The implementation of measures such as simplifying topology and routing, boosting security, and strengthening protocols may help to improve network resilience. |
| Yang et al. [25] | 2021 | This includes an AUV categorization system that takes into consideration performance, formation control, and communication capacities among other factors. This work discussed a complete categorization technique that can be employed across the board will benefit from the proposed paradigm. Additionally, engineers may use this tool to analyse different processes and assist in the selection of the most appropriate forming procedures for a broad variety of applications and circumstances. | ✗ | ✓ | ✓ | ✗ | ✗ | ✗ | ✓ | ✓ | ✗ | S | AUVs with biomimetic bodies, underwater gliders, and torpedo-shaped bodies are all classified in accordance with their body shapes. An AUV is made up of a variety of parts and pieces that work together. In terms of AUV formation control, This work put attention to numerous common misconceptions and questionable findings. |
| Uddin et al. [30] | 2021 | This work presents the in-depth survey in blockchain technology for IoT, including cloud-based IoTs and fog-based IoTs. This work also addresses issues such as research gaps, roadblocks, and possible solutions. | ✓ | ✗ | ✗ | ✓ | ✓ | ✓ | ✗ | ✗ | ✓ | S | IoT-based eHealth, smart home and smart vehicle applications that mix Edge and Fog technologies as well as Cloud computing and blockchain technology are covered in detail. It is observed that IoT still has a number of technical and security issues that need to be addressed. |
| Constantinou et al. [41] | 2021 | It is the goal of modelling in the IoT industry to better understand how to optimise the usage of limited resources. Here, this work provide a method for analysing IoT operational parameters that are linked to resource consumption. | ✗ | ✓ | ✓ | ✗ | ✗ | ✗ | ✓ | ✗ | ✗ | I | This work describes an IoT setup that includes underwater sound detection, classification, and transmission of IoT configuration instructions and results, among other things. This work utilized authentic sounds of ROV thrusters and underwater background noise to train a classification model offline, which allowed us to save time. |
| Bhattacharjya et al. [47] | 2021 | Among the proposed EdgeIoUT architecture components are underwater sensors and cluster heads, drones, SDN switches, and data storage, all of which are part of the system's fourth layer. Comparing our proposed solution to the present traditional Software-Defined IoUT is accomplished via the usage of QualiNet 7.1. | ✗ | ✓ | ✓ | ✗ | ✗ | ✗ | ✓ | ✓ | ✗ | I | Power consumption is a major problem in wireless sensor networks. This work planned to build and simulate a four-layer Software-Defined smart Internet of Underwater Things using edge-drones named EdgeIoUT. The envisioned network will employ SDN switches to operate a variety of drones. |

A: Blockchain-based Solution, B: UUD-based Application, C: Underwater IoT network, D: Smart Contract Discussions, E: Cryptography or Security Aspects or Primitives, F: Type of Blockchain, G: Underwater Resource Management, H: UUD control strategies/architecture/framework, I: Consensus Algorithm, J: Survey (S)/ Implementation (I) / Conceptual (C).

## 1.4 Article Preparation

To begin, the authors searched for publications in significant indexing and search databases such as Google Scholar, PubMed, Web of Science and Scopus. After that, they used their own judgement to look through the items they had discovered. Concerning the objectives of this work, a comprehensive survey is prepared by accumulating the state of knowledge on the blockchain, UUDs, UUVs, AUVs, ROVs, and related practices. This comprehensive survey has developed a detailed discussion and summary of observations from various publications published at reputed venues. To define the search process, the main terms associated with the search process include "IoUTs", "UUDs", "AUVs ", "Underwater Networks", "Underwater Surveillance", "Blockchain for Underwater", "Smart Contract for IoUT and IoT", "Consensus Algorithms for IoT and IoUT Networks", and "Framework for Underwater Networks". In the article collection, those articles were discarded, which does not show qualitative and quantitative explanations to Underwater Technology or Blockchain-based practices.

## 1.5 Article Structure

The rest of the paper is organized as follows: Section II presents related work and background study of Blockchain and UUDs. Section III elaborates the need of Blockchain for addressing the potential challenges in UUDs and Section IV examines the blockchain based computational tools and framework for UUDs. We discuss the Blockchain concept design for UUDs in Section VI. Various Blockchain use cases for UUDs are given in Section VI. Section VII describes the integration of Blockchain with other enabling technologies for UUDs. We provide a discussion on open research challenges and future directions in Section VIII and conclude the paper in Section IX.

## 2. STATE OF THE ART OF UUDs AND BLOCKCHAIN IMPLICATIONS

This section explains the state-of-the-art survey over UUD and blockchain. Details are presented as follows.

### 2.1 UUDs

This section presents the state-of-the-art about UUDs research and the challenges for UUD applications that Blockchain could address. Admittedly, drones or AAVs have been widely recognized for their utilization for military purposes. However, as their popularity has grown in the last years, as well as the awareness among those working in business and research, varieties of commercial and civil applications have been explored (e.g., rescuing, monitoring, filming, delivering goods and medical supplies, etc). AAVs are considered a type of robotic vehicle that can transport payloads and fly operations controlled by remotely controlled either stations or autonomously [23]. AAVs can be classified based on various parameters, and the main classifications are based on the flying configuration, wing type, its weight, and flying altitude. For example, considering the flying mechanism, common generic categories include multi-rotor (rotary-wing), fixed-wing, and hybrid (fixed/rotary-wing). The rapid proliferation of AAVs technology has achieved recent deployments in civilian and commercial applications, such as in extreme applications in underwater environments.

In this context, due to technological advancements, adjacent disruptive technologies such as AUVs have gained attention in scientific studies. Drones have allowed scientific discovers of marine animals' behaviour (e.g., sharks) and the conservation of marine ecosystems, helping to reduce or replace critical and dangerous activities for humans [24]. AUVs are submerged underwater vehicles fueled by fuel cells or batteries, controlled by computers assisted by navigation systems [25]. They are equipped with electronic and mechanical components orchestrated in hardware and software. Basic systems of AUVs include navigation systems, sensor systems and energy systems. Specifically, in this study, we are interested in examining a particular type of AUVs: the UUDs.

The UUDs can be considered a specific category of AUVs equipped with the same basic systems. They offer important benefits like robustness, configurability, and flexibility for exploration in critical conditions [26]. Very recently, there has been a growing interest in the utilization of UUDs in areas of scientific and societal interest, such as in industry (e.g., inspection tasks), marine surveillance, military applications, and maritime activities in unknown underwater conditions [24][25][27]. Drones for aquatic applications have become more available as disruptive technologies that might play a significant role in water system control and management [28]. Besides, visual observation of marine animals using multi-rotor and fixed-wing UUD has been the most typical approach in groundwater investigations [27]. These vehicles often have a camera allowing analysts to assess photos that are communicated in real-time or kept in memory.

Interrelated factors influence UUD choice. The main factors can be organized into the following categories. The first category is the factors influencing the type of drone and payload (e.g., survey area, flight time, sensor payload). In a second category are the factors influencing the flight planning (e.g., the network telecommunication available, drone regulations and laws, site-specific conditions, the use of ocean or beach by humans). The third are the depth of the operation, the operation time and water turbidity. The fourth category affecting the flight and data collection performance directly are the weather, wind, waves and visibility conditions [24].

Most importantly, despite the potentials of UUD using, the extant empirical literature shows that there are several barriers and constraints involving the empirical applications. Typical constraints include the lack of knowledge about the precise location of targets, communication failures, marine animals and vegetation occasionally colliding, battery life, time-consuming underwater missions, limited visibility, nocturnal operations and measurements, inaccurate drone placement, among other issues [28].

On the other hand, the Blockchain protocols arise as a promising opportunity to address these significant limitations [25], leading to improve the current needs of communication reliability in UUD tasks in underwater navigation involving radio-frequency, acoustic and optical communication [29] [30] [31]. Comparatively, underwater acoustic communication is the most frequent mechanism of communication used in marine aquatic operations with UUDs. This is because maritime conditions impose limitations that decrease the communication quality. Examples of limitation include communication propagation delay, communication noise, path loss high and Doppler effect [25].

## 2.2 Blockchain

Blockchain technology, which has been predominant in cryptocurrency transactions, has been proven to provide a highly privacy and safety solution for data processing in networks [30] [31]. Blockchain is a distributed and transactional database that allows for the safe storage and processing of data among a vast number of network members [30]. Blockchain is useful to overcome drawbacks in data communications by mitigating malicious activity in the network, ensuring data immutability, transparency, and operational stability. In peer-to-peer networks, Blockchain uses public key infrastructure for the authentication, authorization of entities, and encryption of information [29].

Compared to conventional centralized types of record-keeping technologies, Blockchain provides advantages by offering trust and transparency and operating without failure throughout transaction processing [23]. The Blockchain architecture is composed of six primary levels: network, data, consensus, contract, incentive, and application layers (Fig. 1).

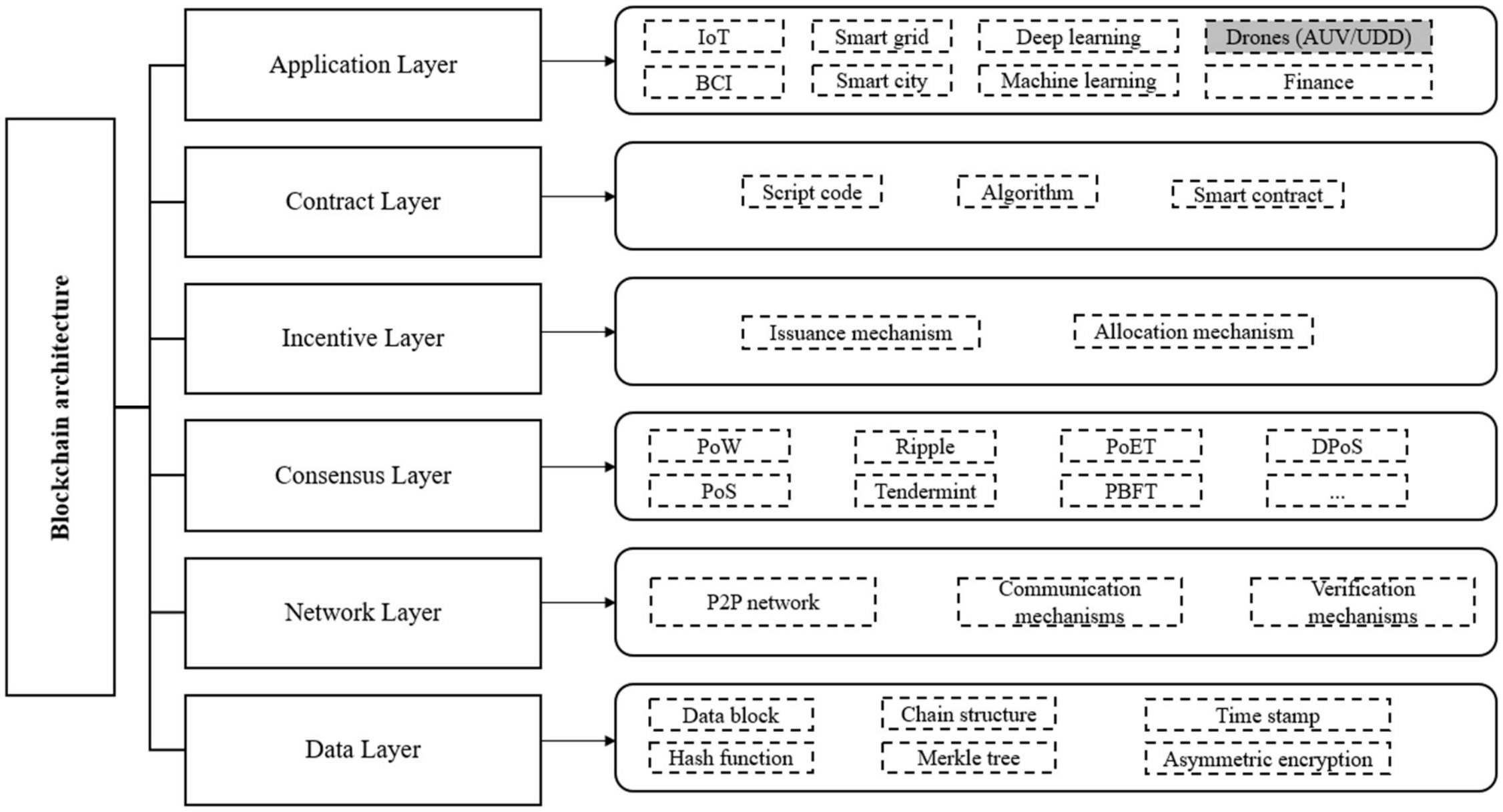

**Figure 1:** Blockchain architecture elements [23, adapted].

The primary data unit of the Blockchain is a transaction, and a number of transactions are structured into a block. The header of each marked block provides the cryptographic hash of its immediate precedent verified block, establishing a sequential link among the data blocks on the Blockchain. This mechanism provides irreversibility, ensuring data integrity [29]. The primary objectives applying Blockchain in UUD as well in broad applications include (i) decentralization, which solves one-point failure problems and bottlenecks in the network; (ii) increased network reliability and security; (iii) improved data transaction traceability; (iv) transparency in transaction records; (v) data privacy and integrity in data transactions or events; and (vi) cost reduction in the system because Blockchain does not require intermediaries [30].

### 2.3 Main Constraints and Challenges for UUD Applications

Several problems involve realistic UUD applications for civil and commercial proposals. For example, when a group of coordinated UUD encounters obstacles in the route, it needs to change the formation shape to prevent the collision. However, the conventional formation control approaches may not be suitable for underwater environments because most of them are based on presumptions that are difficult to implement in underwater environments [25]. Another issue is that the sort of data to be gathered by the UUD will affect (i) the sensors system choice, (ii) the type of payload to be affixed in the UUD, and (iii) the UUD model choice [24].

Furthermore, the technological problems associated with the use of Blockchain in the processing of IoT data [30] are similar to those encountered while implementing IoT for UUD applications. Hence, the following are the current significant challenges (Fig. 2):

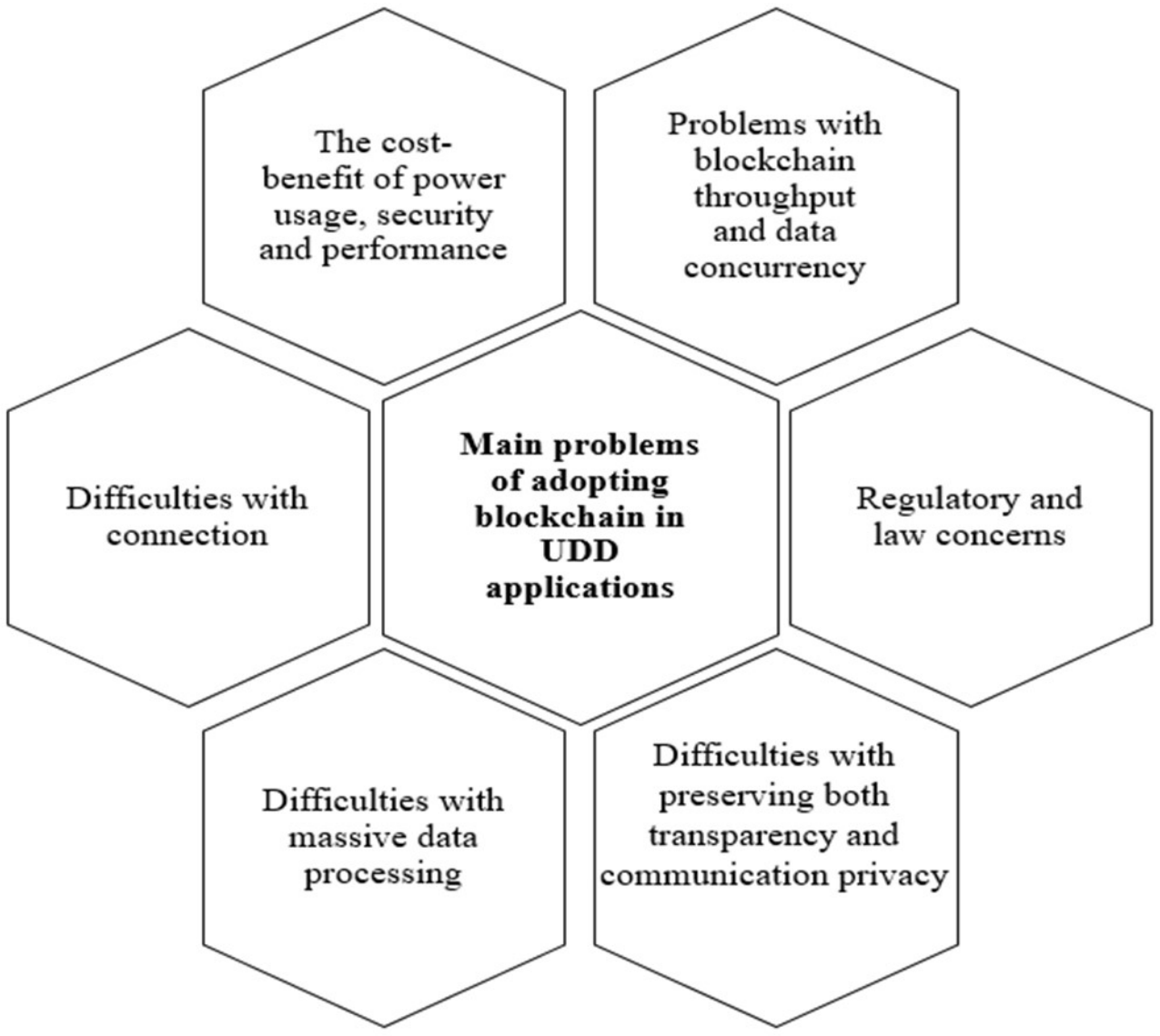

**Figure 2:** Major challenges to adopt Blockchain in UUD applications.

The primary challenge to adopt Blockchain in practical UUD is the cost-benefit of power usage, security and performance. This is related to the idea that the considerable processing power required to execute Blockchain algorithms has inhibited the progress of applications on limited-resource devices. The second problem is related to Blockchain throughput and data concurrency. Because of its sophisticated cryptographic security protocol and consensus procedures, the throughput is restricted. Fast coordination of new blocks across Blockchain nodes necessitates more bandwidth. Third, are the difficulties with the network connection. Fourth, are the difficulties with massive data processing (e.g., originated from UUD swarms) in the Blockchain. Next, there are difficulties with preserving both transparency and communication privacy in the UUD network. There are also regulatory and law concerns, including anonymity, governance, and data records.

### 2.3 Challenges for UUD Applications that Blockchain Could Address

Some assumptions and scenarios considered in drones' operations in the air or on ground conditions are complex challenges for UUD environments [25], where Blockchain adoption can be beneficial. In this case, in the same manner that Blockchain has the potential to be used in UAVs, Blockchain can be used in a variety of applications in UUD systems.

To begin, Blockchain can help with the coordination of UUD services: (i) Blockchain can be used to prevent collisions between individual UUDs and obstacles (e.g., debris or maritime animals) or swarms of UUDs during missions (by storing the latitude and longitude of all UUDs), (ii) to allocate the services load uniformly during the missions, (iii) to authenticate data and entities in the network, and (iv) to enable collaboration and rapid synchronization among UUDs in the network. In this instance, Blockchain-enabled communication channels might be utilized to request help from other UUDs (e.g., in system failures, in sensor failures, low battery levels, etc.) [23] [27] [29].

Blockchain also can be used to enhance the security of UUD networks by (i) reducing interference of wireless signals between the control system and the UUD, (ii) detecting UUD hijacks, altered information, and securing the integrity of the data, (iii) safeguarding UUD communications via legitimate decryption keys, (iv) securing UUD data types (e.g., UUD identifier data, flight route data, sensors data) and data dissemination to end-users. In addition, Blockchain can assist UUD for surveillance applications, (i) keeping high trust in intrusion alerting identification, (ii) safeguarding data from opponents, (iii) lowering external threats to cyber-attacks and physical hazard, (iv) versatile visualize of vast areas, and (v) developing an efficient and secure data transfer framework [29] [30] [31].

Blockchain can contribute to ensuring the legitimacy of data sources in UUDs [25] [29]. Moreover, decentralized Blockchains can be utilized in UUD for data storage produced by heterogeneous aquatic drones deployed in various locations and for validating software update confirmation [26]. The inclusion of Blockchain in frameworks can improve marine applications, particularly those involving cooperation with other robots or UUDs [25] [29]. Lastly, one possible use of Blockchain in UUD applications would be to use the Blockchain in the cloud to perform the consensus procedure for adding the IoUT data to the network [25][32].

## 3. BLOCKCHAIN FOR UUDs

Drones have been used in studying sharks filling knowledge gaps about their behaviour, movement, social interaction and predation across multiple species. Butcher et al. [24] discussed drone technology in shark research and presented a detailed review on current and potential usage of underwater drones in studying these cryptic species. Greengard [33] stated, UUDs are specifically programmed for specific tasks, such as identifying shipwrecks, spotting different types of marine life and executing complex missions. Blockchain being a transparent electronic ledger facilitates uniform service distribution amongst various UUDs, by maintaining load information of different regions, and assigning UUDs accordingly. In case of dynamic load, it uses a mechanism of using non-overlapping coordinates that are randomly generated and assigned to UUDs. Blockchain facilitates the storage of coordinates of each of the UUDs in its database and uses its algorithms on these stored coordinates, to guide optimal routes to the destination, ensuring collision avoidance. With different vendors interacting, maintaining uniform load distribution across various regions is a crucial task.

### 3.1. Security and Privacy challenges in UUDs

Drones and UAVs are effective in specific missions, however, there are several challenges to the security and privacy of data. Small-sized drones, acclaimed widely due to their small wingspan and lightweight body, also suffer from security and privacy issues. UUD communication fabric is vulnerable to threats such as spoofing, DoS, MITM, eavesdropping, and data tampering attacks. Following we discuss potential security, privacy and trust challenges for UUDs networks and how Blockchain can be useful in solving these challenges.

#### 3.1.1. Potential challenges

Lack of appropriate encryption on drone chips lends them susceptible to hijack, and vulnerable to MITM attack. In recent times, the concept of IoD has picked up in defense, industry, military and civilian applications. However, their design is not conceptualized keeping security and privacy issues into consideration. This makes them prone to several issues such as privacy leakage, raising threats to data safety, and allowing for minimal support to data encryption and decryption strategies Drones or UAVs run a danger of manipulation and usage for malicious activities. Attackers can get access to sensitive sensory data of drones, reprogram and misuse it. Moreover, UUDs network challenges are trust and data authentication, maintaining security and reliability of channels and finding optimal paths. Common attacks are unauthorized access to UUD ID and tracing of physical location of UUD, DoS attack, Sybil attack that creates an illusion in UUD network by imitating other UUDs with same ID causing increase in latency of data transmission by spamming, MITM attack, Blackhole attack, sharing fake information of location.

Drone architectural and design issues leave them with vulnerabilities, making way for threats and cyber-attacks. Drones are susceptible to spoofing, with information shared by them spoofed. Attackers tend to use GPS spoofing due to its ease, to get access to the information and manipulate it. Due to vulnerabilities in communication devices of drones, the information shared by drones is susceptible to malware and viral threats. It is prone to data interruption and interception, causing insertion of malicious data during communication between drone and controller device. Drone program instructions manipulated cause them to change their operation and lead to destruction. In event of signal congestion, by using signal loss mechanism, spoofed drones may lead to transfer of control to a third party. Arduino, raspberry-pi microcontrollers prove useful in this. Lack of appropriate encryption on drone chips lends them susceptible to hijack, and vulnerable to MITM attacks.

#### 3.1.2. Blockchain-based solutions

Blockchain technology is proposed for effective authorization and access control mechanisms in drone and drone network security. Blockchain, best described as a distributed ledger, protects shared data by use of public key encryption and hash functions, ensuring trust of information stored and improvement in privacy, security and transparency in UUD. Alladi et al. [23] presented Blockchain applications in underwater drones encompassing secure transfer of data over network, management of inventory, and decentralized storage. They highlighted the challenges of Blockchain integration in underwater drones. In Blockchain, decentralized automation is provided through smart contracts, which have predefined rules sets to govern interaction between interacting parties and help to verify, facilitate and enforce negotiation of transactions in context. Blockchain provides automated data storage, verification, transactions and decision-making. Pham et al. [26] presented a framework to study coordination of multiple underwater drones. The framework is based on object-oriented concepts, implemented in real time, in an open-source environment, for maximum flexibility. Object Orientation facilitates quick development and incorporation of dynamically changing requirements. To avoid MITM, Blockchain uses consensus mechanisms to detect fake information in the network.

One of the ways to handle security issues during communication of underwater vehicles with base stations is through intrusion detection methods of various types, such as rule-based, signature-based and anomaly-detection based. These capture network flow, analyse anomalies, and detect abnormal patterns and spot unlawful activities. Apart from intrusion detection methods, use of forensic methods for monitoring the network flow, is common. Drone data transformed to packets and its communication through the communication channels with cipher security, as a technique proves handy in protecting it from attacks.

Through Blockchain, UUDs can uniquely sign (digital signature) data collected using a private key and broadcast to the whole network. This ensures truthfulness, as data source and entity authentication amongst interacting UUDs, is uniquely identifiable. UUDs in emergencies, seek help from other UUDs in the network, such as in situations, when there are battery issues, faults and malfunction of sensors. Decision histories can be stored in ledgers, and accessed when needed. This facilitates distributed decision-making. It helps to synchronize with other UUDs in the fleet, instead of using processing power and training time. This finds application in military and multi-terrain environments where different UUDs may seek to coordinate and operate. In addition, Blockchain facilitates cooperation between different UUDs that are working in the same environment. These UUDs may be from various agencies that might be sharing competitor relationships at times. They can share common communication channels and confidential data within the network. In practical scenarios, to ascertain which UUD has more control over Blockchain in UUD Networks, mechanisms of weight assignment to it are used.

### 3.2. Routing and Advanced Communication Challenges in UUDs

During underwater communication, the information is required to be sent from the source node to the sink node. The role of routing protocols in underwater communication is highly important to ensure a secure and optimal path between the two nodes [34]. Following, we briefly discuss the routing challenges in UUDs and how Blockchain can be useful to provide potential solutions.

#### 3.2.1. Potential challenges

Due to the restricted resources for each sensor and node during underwater communication, various relay nodes are used for coordination while transmitting the data from the source to the sink node [16]. The design of routing protocols in UUDs suffers from multiple obstacles, for example, acoustic signals take low bandwidth and longer and inconstant propagation delays, i.e., the end-to-end latency for communications for underwater things is longer [8]. Moreover, the energy or power dissipation requirements needed for various sensors and devices are higher and the possibilities of increased error rate and noise are much more during underwater communications [34]. Usually, underwater wireless networks can be separated into smaller regions or clusters. A cluster head is selected from each of the clusters based on the higher resources, e.g., with more energy capabilities. The cluster heads are assigned the major role of routing and forwarding the data from one cluster to another [18],[34]. However, the cluster heads can be targeted by adversaries by launching various attacks causing serious privacy threats [18][36].

Network topologies during underwater communication may vary more often due to the water current among other factors in the ocean environment [9]. With the dynamic change in the network topology, designing the routing protocol for underwater networks is not very straight-forward. Moreover, the localization of the various sensors/devices as well as underwater vehicles and drones will be a challenging task [37]. The conventional methods of localization such as GPS might not be well-fit inside or at the depth of the sea because of the impairments of radio frequency waves [5]. Therefore, the localization-based routing mechanism can also be a massive problem for underwater things. The situation of void holes can also affect the underwater routing mechanism where the information cannot be transmitted from the forwarding node to the next node [38]. Other routing and communication challenges include [37][38]: (i) The bandwidth of acoustic channels is very limited because radio waves cannot be used for underwater communication. As a result of its narrower bandwidth, acoustic transmission requires a greater amount of power to send the same quantity of data as other forms of communication, (ii) As the batteries in UUVs cannot be swapped out easily, there is always a pressing need for low-power communication methods that extend the network's useful life. The quantity of data sent across the network may be reduced to achieve this goal. To achieve this goal, it may be feasible to lessen the amount of energy required for transmission. Like its terrestrial forebears, acoustic channels need a significant amount of power to reliably transmit data, (iii) limited storage and processing capabilities, (iv) Failures in sensors used in UUVs are commonplace due to the fouling and corrosion that may occur in these environments. This may fails in network communication, and (v) Due to the delay in a connection's ability to send a packet, there may be interruptions in the flow of data.

### 3.2.2. Blockchain-based solutions

Number of traditional static routing protocols (such as the reactive routing protocols) are proposed in the state-of-the-art to address the above-highlighted routing challenges. These reactive protocols may use different routing approaches depending on the need of the application i.e., clustering, multi-hop, cooperative and non-cooperative [17]. For example, the work in [14] explored several existing reactive and proactive routing protocols for underwater sensor networks. However, these approaches might not be feasible and appropriate for underwater communication because the location of the devices/sensors may change constantly due to the mobility of water current and thus the network topology in such unstable conditions is mostly dynamic [16]. On the other hand, authors in [138] developed the Q-learning-based energy-efficient routing protocol which includes both reactive and proactive approaches. According to their simulation result, while increasing the network throughput, energy consumption increases.

Blockchain can be useful for ensuring secure and trusted lightweight data processing, sharing and storing in a routing mechanism. Two nodes on the Blockchain network can share the data/information with each other without requiring the trusted third-party. For example, the research work in [18] presented a lightweight blockchain framework for IoUTs which is based on the lightweight cluster-based proactive routing protocol. The proposed network architecture is based on hierarchical monitoring topology which gathers the data at different levels and then transmits it via the lightweight blockchain. Another research in [12] developed a blockchain-based energy-efficient routing mechanism that can avoid the situation of void holes and consume lesser energy. When figuring out how to help mobile autonomous systems plan their routes in a dynamic setting, there are a few potential approaches to go[57]. It took the use of a method that must ensure to programme an autonomous UUV to travel the path that would use the least amount of power. A blockchain-based path travel scheme can guide an autonomous UUV to an adaptable evolutionary algorithm, which can be described to create routes for a large maritime area under practical time limitations. This tactic seeks to identify potential ways of entry into the easy and secure route identification. Those who are employed in the field of manufacturing equipment for secure communication at depth could discover that using a routing algorithm that is founded on blockchain technology works to their benefit. This is because using this technique reduces the amount of time needed to send data between autonomous UUVs in a network. Further, it adds security and feasibility to integrate other advanced technologies. Other blockchain-based solutions include: (i) blockchain ensures efficient and proper use of power for acoustic transmission to send the same quantity of data as other forms of communication, (ii) blockchain can support the timely and necessary integration of advanced technologies (like IoT, cloud computing etc.) for handling the necessary storage and computing capabilities, (iii) Blockchain technology is more accountable in identifying the failures in underwater sensors due to the fouling and corrosion that may occur in these environments more frequently, and (iv) Blockchain network can keep a record of the delay in a connection's ability to send a packet and interruptions in the flow of data. Thus, it is easy to handle retransmission without much delay and repetition of packets.

### 3.3. Resource Management Challenges in UUDs

Since, the underwater communication networks will be a large-scale IoT ecosystem which may contain a massive number of heterogeneous sensors, devices and drones together with needed network and communication infrastructure. Underwater IoT networks usually require higher resources as compared with the conventional (non-underwater) networks, e.g., higher energy or power requirements. Therefore, resource management is a highly crucial research problem in this domain. Following, we explore various resource management challenges in UUDs, and their potential solutions based on Blockchain technology.

#### 3.3.1. Potential challenges

Efficient means of providing resource management has been seen as a vital aspect for underwater communication. Considering the hostile and highly dynamic environment of the ocean, resource constrained network elements, noise and interference during the underwater communication, and higher energy requirements, different secure and optimal ways are required to manage and fulfil the dynamic needs of the UUVs [16], [39], [40]. From the view of communication and networking infrastructure, UUVs will comprise several nodes and devices with restricted network resource capabilities in terms of limited power consumption, processing, storage, and computational capabilities [41]. During different processes in UUVs, multiple nodes may share the communication channel with diverse network elements and the resource demands may vary constantly [42]. Therefore, dynamic resource allocation must be placed in an efficient and optimal manner so that each network element and processes can get the needed resources.

In a shared network resource ecosystem where multiple nodes/entities may require common resources, there may be several challenges in providing the intelligent or autonomous allocation of the resources. Dynamic monitoring and tracking of various available resources are needed to know what resources are occupied and empty and accordingly can utilize them efficiently [15]. Furthermore, it is crucial to ensure that only legitimate and authorized nodes, devices or network entities have the access to the needed resources [12]. Secure and trusted resource sharing capabilities are vital in underwater communication where rich-resource nodes can respond to any request made by the other nodes in the network asking for the specific resources.

#### 3.3.2. Blockchain-based solutions

To address the above-mentioned resource management issues, several resource allocations algorithms were developed in the literature. For example, the research work in [16] proposed a self-adaptive Q-Learning algorithm for distributed resource allocation for underwater communications. In the algorithm, nodes constantly interact with the environment and learn about the optimal allocation methods and condense the interference of the network. Based on multi-agent reinforcement learning, authors in [43] formulated an adaptive and distributed algorithm for managing the resource for IoUTs. However, many of these studies have not much discussed regarding the secure, immutable and trusted resource sharing among various nodes in the network.

The integration of Blockchain technology with the UUDs will allow them to fairly monitor and manage the efficient resource utilization during underwater communications. The distributed digital ledger is shared with each involved stakeholders or network elements that can keep track of the allocated and idle network resources [13]. With this way, it is possible to fulfil the dynamic needs of resources in the network. Furthermore, smart contracts running in the blockchain can be used to enable intelligent decision-making for resource allocations. Blockchain reduces the administrative cost by removing the need of additional intermediates for managing and allocating resources. Blockchain can also offer the secure resource accessibility mechanisms for various network elements, so that only authorized entities can have access to the resources [44].

#### 3.3.3. Data Management Challenges in UUDs

UUDs are expected to gather a huge volume of heterogeneous data from diverse connected sensors/devices. Therefore, UUDs will have challenges in securely handling and managing heaps of the data at various steps during the data acquisition, processing, analytics and storage. Following, we discuss the key issues in the context of data management for UUDs and how Blockchain technology can help in overcoming these challenges.

#### 3.3.4. Potential challenges

There are numerous sources in the UUDs environment where a massive amount of the data can be generated such as sensors, devices, drones, vehicles, and surveillance cameras among others. In addition, the collected data can be of multiple types and the data quality might be poor in some cases [17]. Since, the volume, velocity and variety of the acquired data is vast, it requires sophisticated big data approaches for processing and handling the big data [36], [44][45]. The mechanism of capturing, transmitting, analysing and storing huge volume of underwater information enable the emergence of a new research concept namely 'marine big data' [46]. One of the major obstacles is to do suitable classification of the various data types to extract the useful information.

Another major challenge is regarding the secure storage of enormous data, i.e., where to securely store the sensitive data? Cloud based platforms are usually considered as the default option for storing large volumes of the data. However, for delay critical processes in UUVs, part (some) of the data can be stored and processed at the edge or fog network [47], i.e., closer to the vicinity of the source. In addition, due to the unstable network connection of the underwater environment, it is also crucial to have some of the data processing and storage locally (on nodes/devices). Secure and trusted data sharing among various nodes/devices and network elements is a vital requirement during underwater communication [12].

### 3.3.5. Blockchain-based solutions

Blockchain as a distributed ledger technology can provide efficient solutions to data management challenges. For example, Blockchain can enable a tamper-proof computing environment for data storage as well as data sharing among various entities. Distributed ledger can take care of secure data processing, storage, and trusted data sharing among various nodes. Blockchain technology empowers automated data sharing through the agreement made through the smart contract. The integration of Blockchain with edge computing can be done to provide several benefits of blockchain to the edge network, e.g., trusted data sharing among edge nodes. The work in [48] developed Blockchain and edge-based secure storage management for IoT networks. Authors in [49] formulated a trusted data management mechanism with the fusion of edge and Blockchain. Table 2 shows the analysis of recent work conducted in blockchain integrated approaches and limitations for drone-based systems.

Table 2: Various Blockchain integration approaches and limitations for drone or IoT-based systems.

| **Authors** | **Year** | **Approach & Key strength** | **Limitation** |
|---|---|---|---|
| Pham et al. [26] | 2019 | Multi drone coordination studies, specific to formation control accomplished; Object-oriented methodology and its advantages in terms of managing future advancements and changes discussed; | Blockchain integration and application to multi drone environment and formation control problems not discussed |
| Uddin et al. [29] | 2019 | A multi-sensor monitoring hierarchical architecture, with customized blockchain technology integration, for secure routing of IoUT sensor data proposed. Simulation results are presented for efficacy of the approach. | Focus is limited to secure data routing over hierarchical topology within the proposed architecture |
| Rani et al. [50] | 2016 | A hacking procedure on a UAV demonstrated with results highlighting the unauthorized access to sensitive data, fraudulent use and undesirable and irreversible damage | Focus is only on cyberattack demonstration on communication link between UAV and operator, to alter flight path. Methods to mitigate and handle are not elaborated |
| Yaacoub et al. [51] | 2020 | Use of drones for malicious purposes by exploiting vulnerabilities such as communication links and other cyberattacks are reviewed. Detection methods are discussed. Counter measures are suggested. | Blockchain as a solution to cyberattacks is not applied to discussed attack scenarios |
| Pham et al. [52] | 2020 | Formation control problem of multiple low-cost drones, is approached through a leader follower architecture, with DANNC implemented. Suggestive configurations to shape the group of drones, to cross over from narrow areas. Evaluation is done in experimental scenarios | Focus is on coordination control for multi drone formations, secure data transfer issues and Blockchain integration is not discussed |
| Strobel et al. [53] | 2020 | An approach that exploits blockchain to achieve consensus in robotic swarms, in spite of the presence of byzantine robots is presented. To facilitate future research the approach is released as open-source software | While applicability in robotic swarms for secure financial transactions, by exploiting blockchain, is discussed, relevant scenarios have not been elaborated |
| Santos De Campos et al.[54] | 2021 | A blockchain based surveillance framework for multi drone coordination and secure financial transactions is demonstrated. Mechanism of subscription of surveillance services through tokens is detailed. IOTA blockchain is recommended, and highlighted decentralized decision making and coordination strategy with strength of being computationally light. | Comparison of framework with other contemporary works on multi-drone coordination and formation control problems is not discussed. |

## 4. COMPUTATIONAL INVESTIGATIONS, TOOLS, FRAMEWORKS AND MODELs of FIXED WING UUDs

Recent attempts to integrate Blockchain in UUD into the practical application entail further investigation into computational methods, tools, and frameworks for UUD. For example, recently, findings of multiple UUD applications (e.g., lakes, rivers, urban canals, waterways, locks) integrated with sensors and video cameras in the Netherlands are reported by de Lima et al. [28]. The study found that UUD enables data collection that would be cost-prohibitive or even impossible to reach using other approaches. For example, they reported that UUD could map different locations with different vegetation, can obtain 3D data and underwater pictures, establish links between wildlife species with local water quality, and monitor variations in water quality metrics. Nonetheless, a number of technological limitations are outlined in the current literature, such as the integration among systems to make the UUD operations more feasible in practice.

There is a growing interest in control cooperation of swarms of UUDs [26]. A recent study M. Khan et al. , for example, proposed an AUV-assisted energy-efficient cluster control technique for IoUT applications in deep-sea underwater wireless sensor networks. In the framework tested, the AUV in the cluster collects data from other clusters while helps in the cluster head selection and cluster formation.[55]. Other interesting work in the thematic of UUD cluster formation and coordination, found that UUD could be used in large-scale underwater acoustic networks to improve the network reliability pursuing different functions in coordinated groups of UUD and in UUD-assisted networks, where the information is collected from stationary nodes and relied to surface [25].

Already the work developed by Babatunde et al. [27] created a UAV system outfitted with a maritime acoustic recorder to aid in the observation of harbour porpoises in a UK program for environmental monitoring and conservation in oceans. The primary outcomes have shown that the system is able to make autonomous navigation, continuous landing, take-off, and data gathering.

Recent research Pham et al. conducted a simulation-based investigation to evaluate a framework for multiple UUD coordination, especially formation control, in an open-source environment utilizing real-time object-oriented concepts. On the whole, we observe that the outcomes from the literature also suggest that more dedicated simulation systems incorporating UUD and Blockchain features and configurations should be developed. This is due to the complexity of integrating and implementing Blockchain in drone UUD systems, requiring extensive testing prior to practical use [23][26].

Uddin et al. developed a sensor monitoring system with several levels for underwater applications. The system tested comprises fog and cloud components to safely process and store the IoUT data using Blockchain. The literature notices that Blockchain-enabled distributed systems have evolved to enable underwater IoT data to be saved safely and economically without depending of intermediate trusted authority while preserving network privacy via the use of the consensus process. The consensus mechanism, in this case, happens when parties in the Blockchain network engage in processing and assessing IoUT data before certifying IoUT data insertion upon on network [29].

In this research, we presented a comprehensive overview of current scholarship on utilizing Blockchain in UUD applications. The combination of the factors involved in the use of UUD and Blockchain showcases the difficulties faced by scholars and practitioners in realistic implementations. This implies that studies developing computational testing, summation models and new tools and frameworks UUD are required to incorporate Blockchain and UUD configurations and features to assist the decision-making in projects.

Specifically, this study looks at the basic background and technical issues that researchers should ponder while working with UUD and how Blockchain can address the primary challenges of UUD applications. This research also identified knowledge gaps between theoretical research and real Blockchain implementations in realistic UUD operations. By examining the current progress of the area, our findings allow to affirm that the research in this theme is still in its preliminary stages of understanding.

## 5. BLOCKCHAIN CONCEPTUAL DESIGNS FOR UUDs

It has been observed that no or least efforts are made in integrating Blockchain with UUDs. Figure 3 shows an example of incorporating some subsystems of UUDs and Blockchain concepts. Here, 12 Blockchain nodes present the different subsystems (Lighting, Cameras, Incandescent, AC/DC supply, Sensors, High intensity discharge,

Fluorescent, Power Source, Manipulator, and Tool, LED, Tether, and Underwater Connectors) of a remotely operated drone. Each node can have one or more Blockchains storing subsystem events and data. This data will be immutable, secure, transparent, and distributed across all nodes. As a result, subsystems are connected physically, logically, and using Blockchain. Thus it optimizes the vehicle's performance and usages in various applications. Figure 4 shows various UUD and Blockchain concepts that are studied independently. However, the integration of two technologies should have an in-depth analysis of these concepts and their proposals.

Likewise, various advantages of Blockchain technology can be taken by integrating it with underwater drones. Few of the recent underwater drone studies and how Blockchain can be integrated with those studies are briefly explained. Deutsch et al. [56] established a methodology for evaluating the qualitative performance of underwater vehicles and investigated the transit performance of the most sophisticated versions. The transit performance of Slocum, Spray, and Seaglider gliders, as well as propeller-modified variants of these types, is evaluated using a simple glide measure. This study compares and contrasts underwater gliders powered by propellers with underwater vehicles propelled by propellers. The findings show that while operating under ideal conditions, gliding locomotion is more efficient for the various hull shapes. This data can be securely stored in the Blockchain network. The glide metric information stored in the Blockchain network will be strongly protected with cryptography primitives and protocols, generate a real-time response with smart contracts, and integrate consensus algorithms for efficient network building and data sharing. On the other hand, biofouling conditions lead glider performance to incur a twofold penalty, causing gliders to perform worse than propeller-driven vehicles in certain circumstances. A flight dynamics model was built to predict qualitative performance, to test if the Slocum data set described it accurately. For design optimization, even semi-empirical and analytical models are proven to need complex parameterization. If the computational efficiency of the model proves to be helpful for design engineers in the early phases of the design process, then the design engineers will find this very advantageous. This study team used the model to investigate the relationship between wingspan and gliding efficiency, and discovered that the Slocum glider design is near its peak efficiency at the moment. There are three types of nodes on the Blockchain network whose advantage can be taken in an UUD: one-half node, general node, and mining nodes. Blockchain technology can interconnect multiple underwater entities (including drones) and consider each entity as a node. Transactions prepared and broadcast over the peer-to-peer network can be created through the half node and general node in an underwater Blockchain network. Further, When miner nodes detect transactions, they build a Merkle tree out of those transactions to create a Block. Miner counts and iterates through nonce to arrive at a hash code for the Block until a goal has been reached. The process of verification has finally concluded. Afterward, all nodes except the half-nodes append the Block to the end of the current ledger. Likewise, Blockchain can be integrated with the internal or external functionalities of underwater drones. Eichhorn et al. [57] identified the importance of underwater drones. Underwater gliders make it possible to monitor the ocean's surface conditions more effectively. Instead of recording oceanographic data at certain depths and from a single location, gliders may log data for up to one year by following pre-determined routes, unlike buoys' limited data recording capabilities. It is possible to calculate the depth-average velocity by combining the horizontal glider velocity and the GPS update, as well as by utilizing data recorded by sensors that measure depth, such as a conductivity-temperature-depth sensor, in addition to the data recorded by the horizontal glider velocity and GPS update. Navigating or planning glider missions may also be accomplished via the use of horizontal velocity. Underwater drone functionalities, observations, and associated statistics can be stored in the Blockchain network to ensure high security, lesser chance of data loss, streamline data accounting, analysis, and predictions, and reduce operating costs. Eichhorn et al. [57] provided the findings of an investigation into the most accurate way to estimate the horizontal glider velocity possible. Slocum glider flying models that have been used in practice are presented and compared to do this. This study's findings have been used for the development and clarification of a glider model, which explains how steady-state gliding motion operates. To get different model parameters, the nonlinear regression technique are used, and the methodology is further described. Both a descriptive explanation of the criteria for recorded vehicle data and a robust method for correctly calculating the angle of attack are described in this article. Blockchain technology may help ensure the safety of personal data in this instance. Data gathered on flights that utilized gliders over the Indian Ocean was used to validate the procedures presented here. The research demonstrated that a time-varying model is needed to achieve a satisfactory fit between recorded and simulated data. The recorded and simulated parameters stored in Blockchain, if made possible, will make data immutable. Thus, comparative analysis will be more trustworthy, in addition to biofouling. When organisms attach themselves to and grow upon the glider's surface, other factors contribute to the changes. It may be feasible to improve the dead-reckoning algorithm, and the practical use of the proposed method for reading an appropriate horizontal glider velocity may be applied to the glider's process for calculating the depth-averaged velocity if the technique is successful. Comparing various ocean current models with glider-logged data is being done using the depth-average velocity, calculated using the depth-average velocity. If stored in the Blockchain network, all of these statistics provide a more reliable, better performance, secure, and distributed data storage system. The subsequent sections explain the Blockchain integration feasibilities with different parts of UUDs. Details are presented as follows.

### 5.1. Blockchain and UUDs Composite Components

In an UUDs system, there are many composite components, including an engine (consisting of the nose cone, guide vane, main engine, electronics housing, exhaust flap, inlet housing, inlet guide vane), sensors, a tether, a tether management system, a flotation pack, and a thruster. The vast majority of UUDs are equipped with sensors for video cameras and lighting. Additionally, sonars, magnetometers, water samplers, and sound velocity measuring devices may be present. Additionally, ROVs may have a 3D camera installed to enhance the pilot's understanding of the underwater environment. Blockchain technology can be integrated with IoT and sensors associated with various composite components. Blockchain can securely store data collected from IoT and sensor devices. After that, this data is made available in a distributed ledger system. Blockchain can ensure transaction execution as per service or requirement i.e. payment to owner of the underwater drone can be made based on its functionality and usages. This section discusses the features and functionalities of three important UUD parts. Further, the importance of Blockchain in subsystems and to the whole UUD is explained. Details are presented as follows.

- **Wingspan, Wing Surface Area, and Importance of Blockchain:** Wings are very critical to UUV design. In any drone, the wing plays a significant function in providing lift. An example of inspiration from the bird is the twin tabard wing of the tropicbird. This particular wing is situated over the other. To counteract the pull of gravity, a huge fixed-wing aircraft is built with a high lift capacity, making the UUV able to float. Fixed wings, morphing wings, changeable wings, and Quadcopter constitute the standard wing configuration for aquatic UUVs. UUV (fixed-wing) aircraft can take off and land on water, but they cannot go underwater. The fixed-wing UUV is navigated using the technique of surface wave navigation and flight. Engine power will be turned off when in the wave effect. This design uses lightweight wings while maintaining neutral buoyancy in water. Due to its lifting surface, it is more resistant to fatigue. One of the flying fish characteristics is its transition wing with an electric motor and a single propeller. The wing's morphing ability enables it to fold its wings for enhanced underwater movement. The important forces that make the underwater drone operate include diving, rising, drag, thrust, or propulsion. Blockchain technology associated with an UUD can automate the drone operation with more efficiency. Blockchain technology, with the help of IoT and sensor technologies, can control the drone remotely. It can vary the pressure at the top or bottom of wings to control the movements. Further, Blockchain's distributed ledger stores information collected from wings. Figure 5 shows an example of Blockchain technology associated with wing designs. There are multiple subsystems associated with wing designs, including wing-based underwater drone-detection, underwater drone movement and control, wing fabrics, win pressure control systems, and wing forward and backward swept systems. Drones can be identified with object detection methodology in wing-based underwater drone detection after applying AI or ML algorithms. In an underwater drone movement and control system, pressure around the wings can estimate the drone movement. Thus, information on drone movement can be collected or controlled. Likewise, wing fabrics, wing pressure control systems, and wing forward and backward swept systems can give different information about drones and their wings. This information can be stored in private, public, or consortium Blockchain networks. The advantages of Blockchain networks include data immutability, transparency, enhanced security, and automation. Ranganathan et al. [58] stated that the idea of "change of volume" is used to create a cost-effective and straightforward underwater glider. A prototype RoBuoy has been developed using the smallest pieces possible while keeping all of the critical components hidden. Using mathematical modeling, the system has been optimized in size and dimensions to improve its gliding performance to the greatest extent possible. It has been successfully tested experimentally to a depth of 5 m with the same results. It is also feasible to do further performance tests on this glider depending on the wing position and hybridizing it. Zihao et al. [59] describe the hydrodynamic form design and optimization of a low-cost, HFWUG. Flight characteristics of the HFWUG are also examined. Zihao et al. [59] have investigated the effects of various flying wing design factors on the lift-drag ratio, intending to increase the glide efficiency to the greatest extent feasible. Comparing the final design of the HFWUG to the historical gliders, which have an average L/D of 5, the final design of the HFWUG has an L/D of 13.8, which is higher. Performance prediction and simulation in the vertical plane are performed based on this model. The results show the motion parameters in different motion states and the glide trajectory in the vertical plane. All indications indicate that the flying wing design will improve the performance of underwater gliders, making them more suited for gliding in a wide area of application and at a modest glide-slope angle. This has also been tested to a depth of 5 m in the laboratory, and the results were the same. Additional performance experiments may be conducted on this glider if the wing location is varied and it is hybridized. The design and optimization of a cheap, HFWUG with a wingspan of less than one meter are described in the study by Zihao et al. [59]. The flying characteristics of the HFWUG are further examined. Zihao et al. [59] studied the various design factors and concluded that improving the glide efficiency to the greatest degree feasible was achievable by using a flying wing design. The HFWUG design has an L/D ratio of 13.8, whereas earlier gliders had an average L/D ratio of 5. This model shows that real-time performance prediction and simulation are carried out using this design. The research indicates the motion parameters and glide trajectory while in different motion states, as well as the plane of the results in the vertical plane. When designing the flying wing, it is very likely that the aircraft's overall performance will be improved, which will allow underwater gliders to glide over a broader variety of terrain and at a shallower glide-slope angle. However, it is more essential than the interior area to be flat and wide to enhance the length-to-width ratio. This is because the airfoil's maximum thickness cannot be more than the L/D ratio of the airfoil. To accommodate payloads and equipment, there is sufficient space on both sides of the wing. This investigation will serve as the basis for the creation of a prototype for the HFWUG. Appendages to the glider, such as winglets and flaps that may reduce drag while improving maneuverability, will be researched and further enhanced.

- **Chord Length and Importance of Blockchain**: Sadraey [60] discussed the detail on the form of the wing and the length of the chord. Historically, the four-digit NACA airfoil sections have been the most commonly used and the most straightforward to fabricate in the industry. When two parabolas meet at the intersection of a four-digit airfoil, a camber is generated. An initial parabola is traced from the trailing edge to the maximum camber, and this results in the creation of a camber geometry. It is the result of a modified parabola that gives birth to the camber form, which starts at the trailing edge and continues throughout the whole camber shape. The maximum camber of a four-digit NACA airfoil is expressed as a percentage of the chord of the airfoil. To calculate the maximum camber position to the chord length, one may use the second value as a reference. The thickness to chord ratio is shown by the final two digits of this measurement. A zero in the first number of a section of an airfoil indicates that the airfoil is symmetrical. The maximum camber for the NACA 1408 airfoil section is ten percent, while the maximum t/c is eight percent for this section (with two digits after the decimal point). The camber is at its greatest when the chord length is 40 percent of the total length. The drag generated by these airfoils is much greater than that of the preceding generation.
- **Blockchain and Fuselage Length Designs**: Designing fuselages is a difficult task for underwater drones with fixed wings. When the fuselage pitches concerning the wing, it creates pitch-plunge. A long fuselage may introduce inertial coupling when the drone's roll rate influences the pitch. Fuselages, meanwhile, contribute to the overall stresses imposed on the pitch effectors. Additionally, when the angle of attack is high, the existence of vortices created by the fuselage, especially concerning flat or curved bodies, may result in pitch break instability. The fuselage's geometric design is another important component of the design. Due to a few key variables (such as the width-to-height ratio and cross-section form), the shape of the fuselage is determined. Additional information about the item may be obtained using these geometric measurements and shape characteristics, such as the surface area and interior volume. All of this data, if stored in Blockchain, will be useful in various aspects. Using an AI-based object detection and Blockchain network, the data, including pitch-plunge, fuselage geometry, and distribution of width and height information, is helpful to monitor the underwater drone. Further, if data stored in a Blockchain network is compared with real-time values, these characteristics will benefit underwater drone detection in performance and usage optimization. Figure 6 shows how fuselage design and Blockchain-based data storage and access system can take advantage of each other to optimize the UUD performance and security. There are various concepts in fuselage design that can share drone or vehicle data through sensors or object detection-based approaches. This data can be shared and stored in the Blockchain network in different ways. For example, data can be stored in permission or permissionless Blockchain systems. Smart contracts can be designed for transaction defining and invoking associated with data.

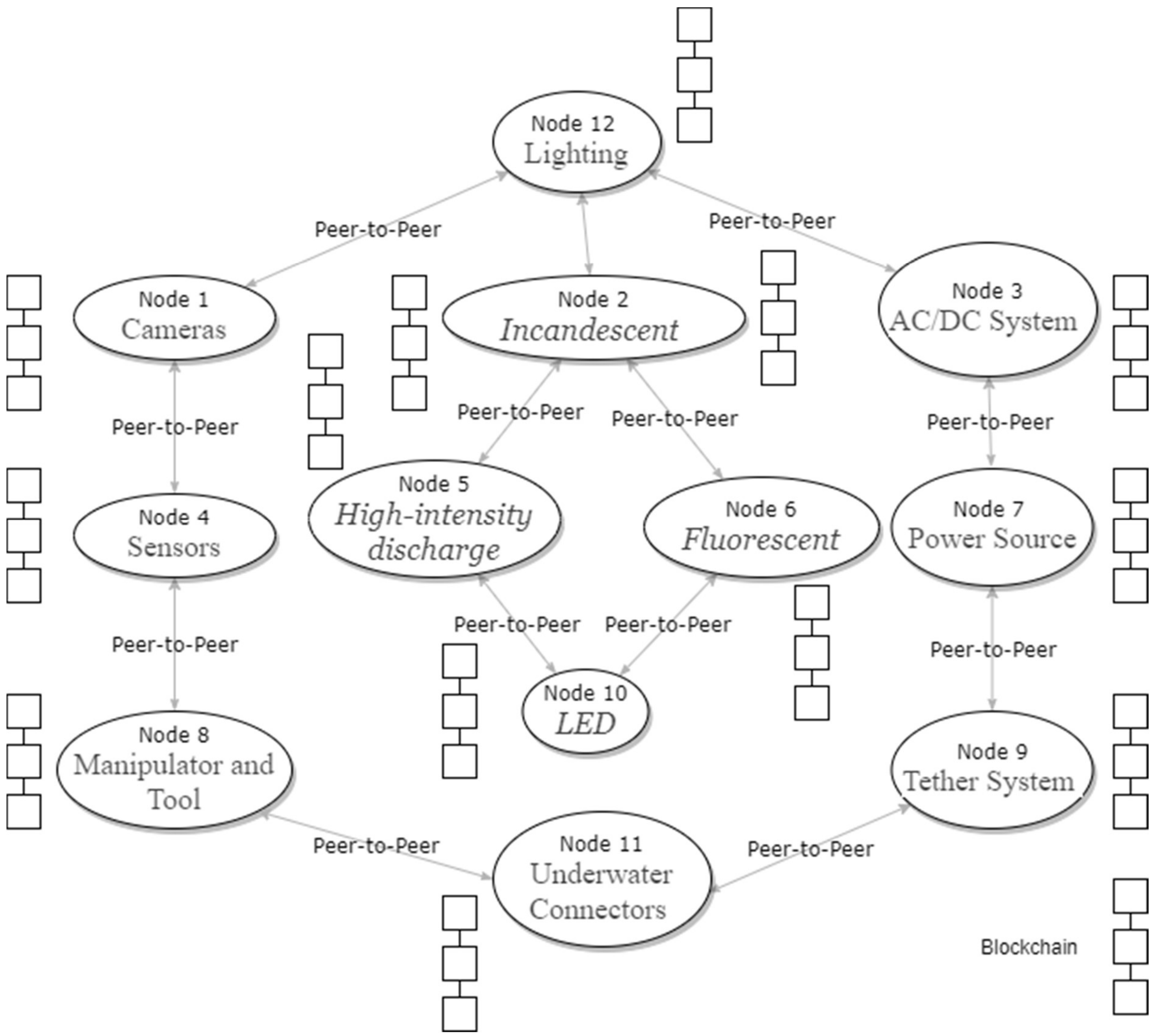

**Figure 3:** An example of remotely operated underwater vehicle's subsystems' Blockchain.

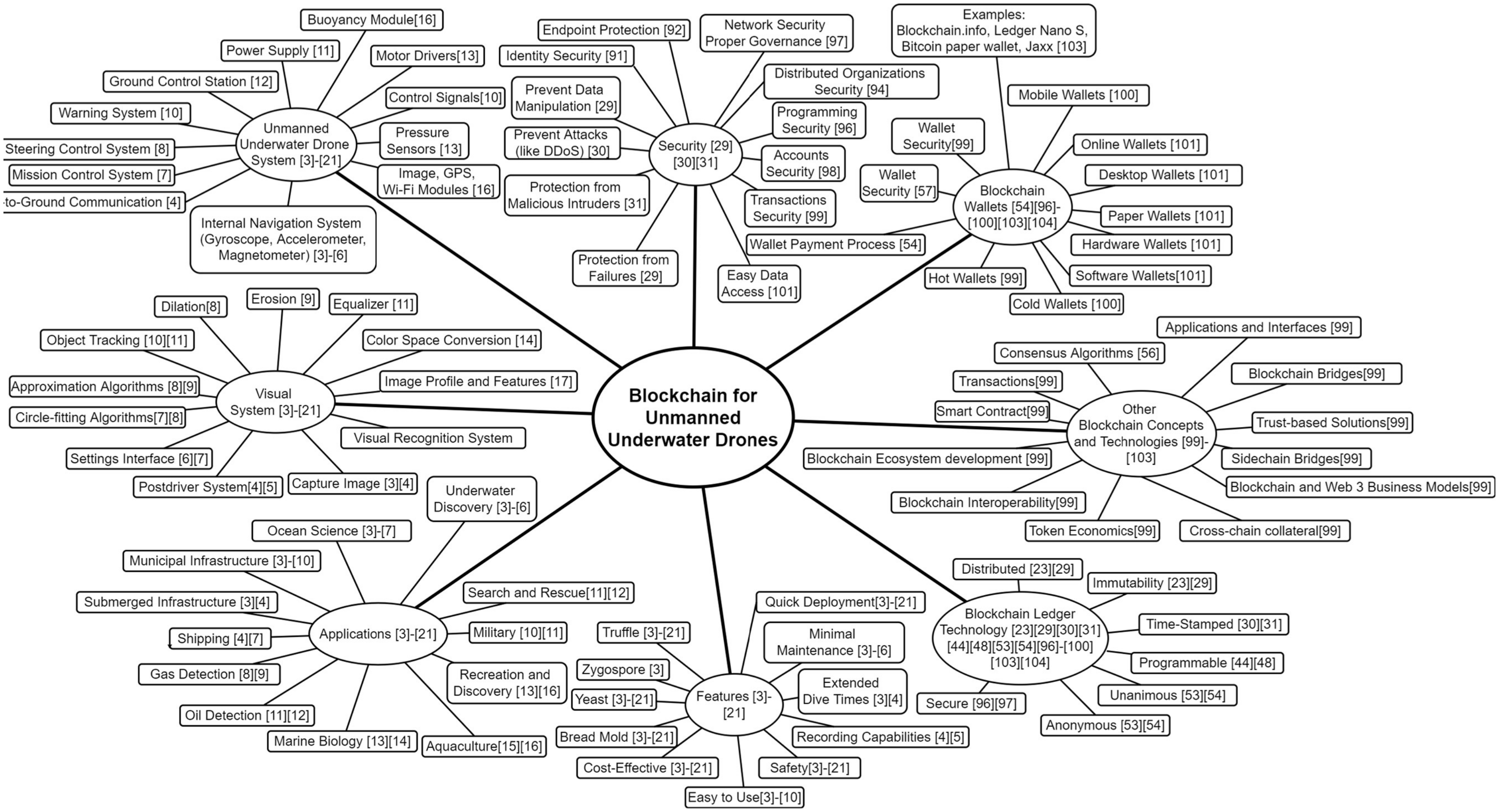

**Figure 4:** Blockchain and Unmanned Underwater Drone

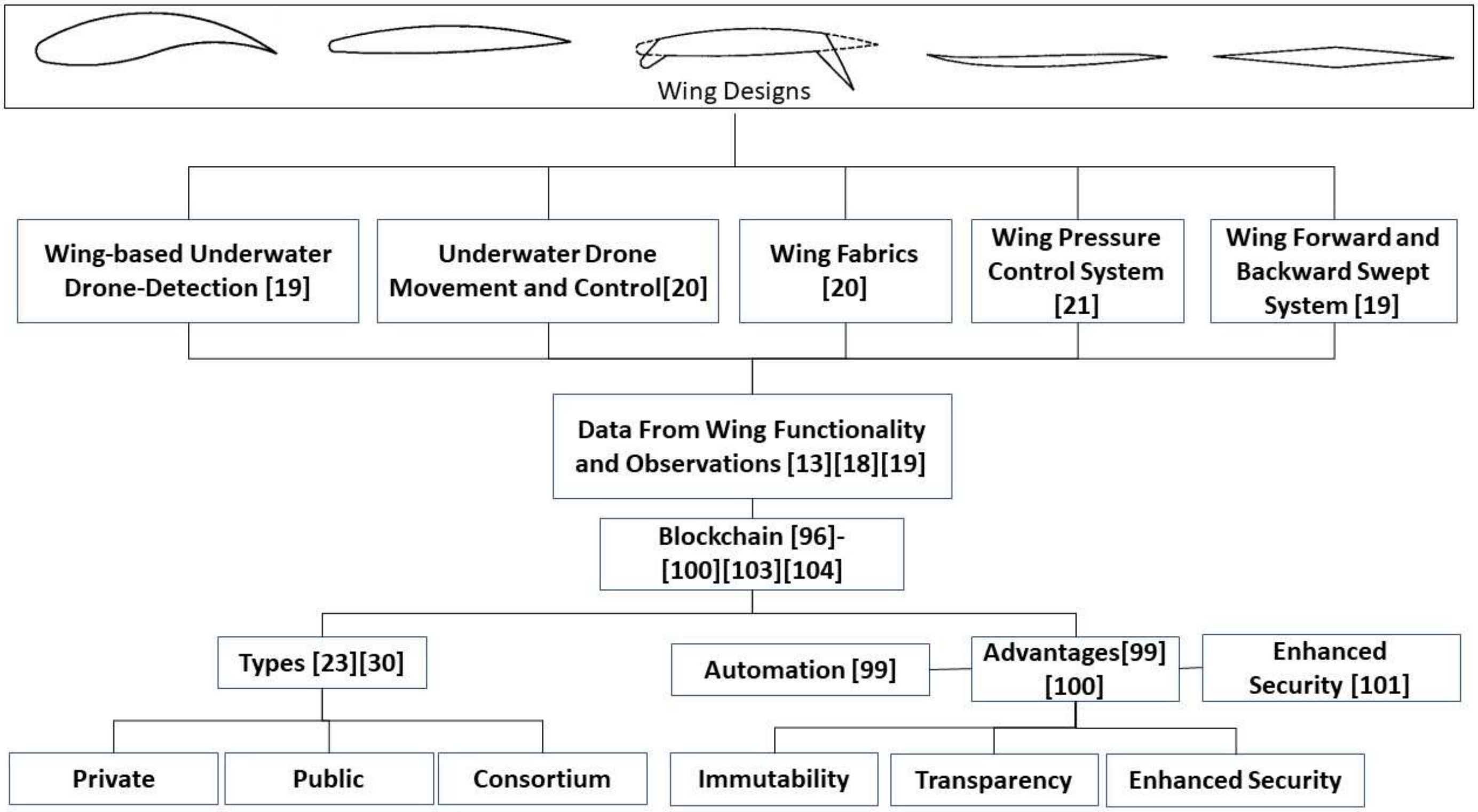

Figure 5: Blockchain and UUD's Wings

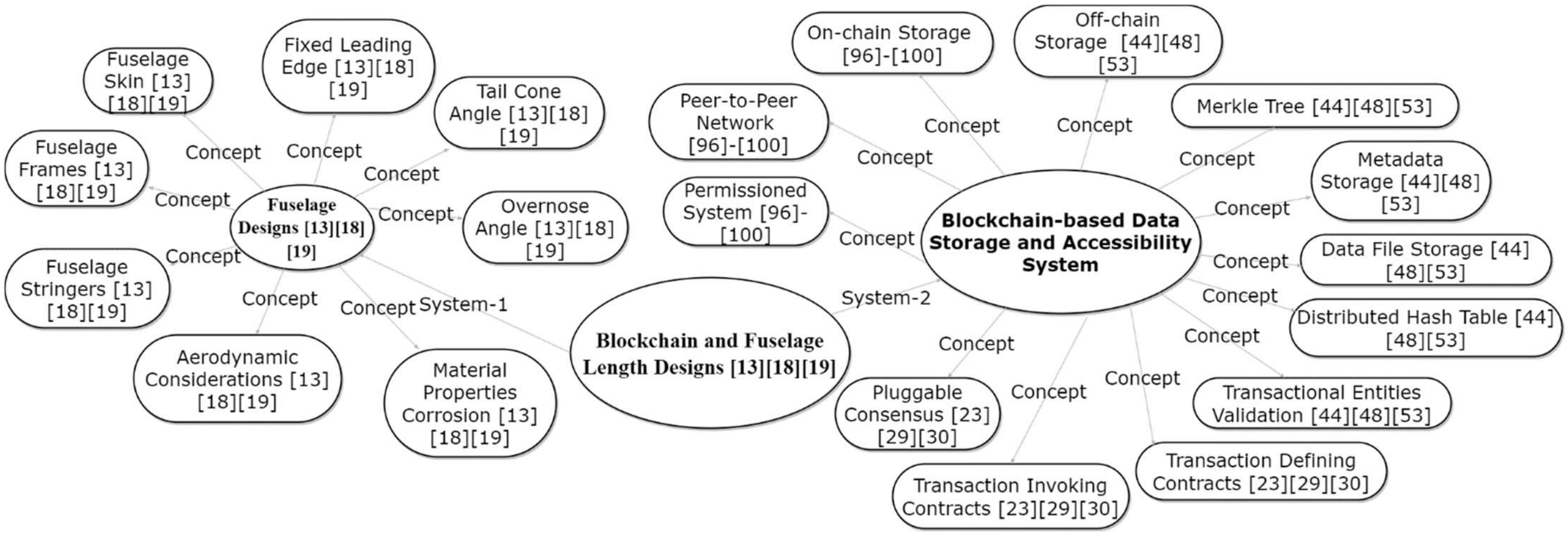

Figure 6: Blockchain and Fuselage Designs

## 5.2. Empennage, Powerplant, Sustainer Engines, and Landing Gear Design, and Importance of Blockchain

The Empennage, Powerplant, and Landing Gear Design are important parts of an underwater drone. The important features of these three parts are [61]:

- Every component of the empennage is included, from the horizontal stabilizer and vertical fin to the elevator and stabilator, including any trim tabs that may be present. The horizontal stabilizer and vertical fin are two examples of permanent surfaces that may be seen in nature. They function similarly to the feathers of an arrow in that they assist the glider to maintain a straight course through the air when flying through the air at a high altitude.

- Glider and engine are other two important parts. Gliders equipped with self-launching engines are powerful enough to lift off without using an external source. The engines may be used to maintain the drone underwater if the circumstances allow it. Some difference exists between how the engine and propeller are situated and what propeller is utilized in self-launching gliders. Motor gliders in which the operator tours or cruises the area and high-performance self-launching gliders are the two types of self-launching gliders one can purchase.
- Sustainer engines, which enable gliders to stay aloft for a prolonged time before being retrieved, are installed on specific gliders. The glider cannot rise off the ground without assistance, even with the use of sustainer engines. Aerotows may be used to launch sailplanes in two ways: either by air or on the ground.
- The main wheel, a front skid, and a tailwheel or skid are all landing gear parts. The landing gear may include both wheel or skid plates as well as wingtip wheels in certain versions. With the assistance of a tailwheel or tail skid, it is usual for gliders to use retractable main landing gear in case of an emergency landing. Breakaway wing tips may be seen on high-performance gliders.

Figure 7 shows the integration of Blockchain and UUD Hardware (Empennage, Powerplant, Sustainer Engines, and Landing Gear Design). In this integration, there are four layers proposed to support the Blockchain features in UUDs. The underwater drone layer considers the drone operations underwater. The Blockchain layer collects data from various sensors. This includes data of empennage, powerplant, sustainer engines, and landing gear design. Here, a node can create one or multiple Blockchains with version, timestamp, file hash, file size, location, Merkle root, and signature information. The bridge layer allows creating a bridge between Blockchain at the Blockchain layer. and individual hardware Blockchain. Here, federated or trustless bridges can be created. The record layer permits the creation of individual hardware Blockchain. Each of the hardware Blockchains contains data associated with hardware. Thus, it allows monitoring, operate and control drones efficiently. UUDs may be programmed to prevent them from getting near critical infrastructure including the underwater electric grid, military equipment, or animals by using the Blockchain. In addition, the technology may also help with industrial inspections and monitoring, and also enhance them.

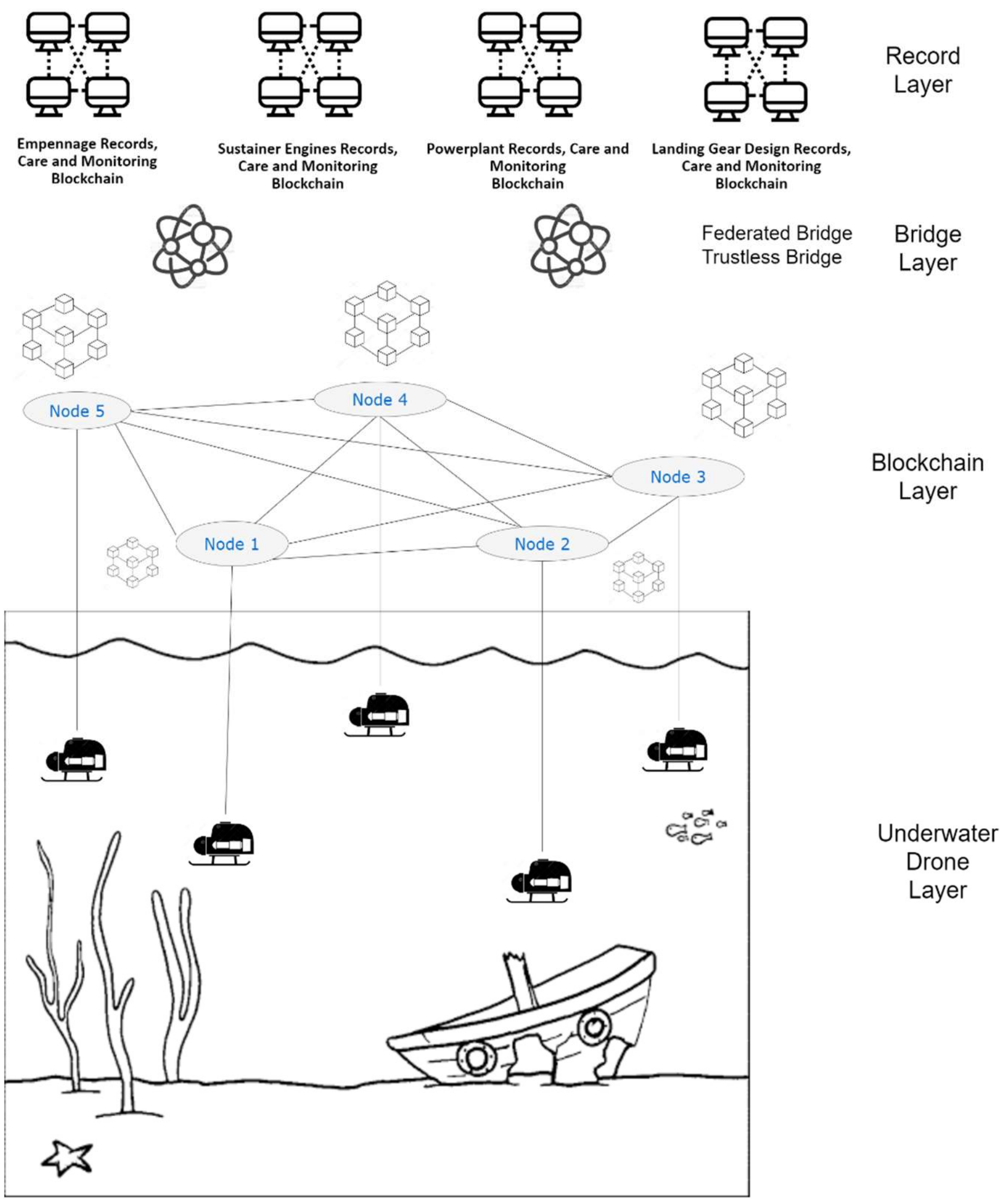

Figure 7: Blockchain and UUD Hardware (Empennage, Powerplant, Sustainer Engines, and Landing Gear Design)

### 5.3. Blockchain and Propulsive System for UUDs

Numerical and experimental techniques have been in use since the dawn of time for all sorts of studies and research. They are still applicable now with regards to marine vehicle propulsion. In research where the Strouhal number and maximum angle of attack are altered, the effects of increasing the Strouhal number and maximum angle of attack are investigated further. Various propulsion systems for underwater vehicles or drones include [62]:

- **Lift-based Propulsion**: Using the flapping foils or wings of a drone, a vehicle such as a drone generates lift and propulsion on the water's surface. As a result of the pressure differential between the upper and bottom surfaces of the flapping wing, a pressure difference is produced. Pairing the wings serves as a means of generating lift and propulsion.
- **Drag-based Propulsion**: Frogs, geckos, and alligators utilize a method that comprises of two movements: first, they use their bodies to push water backward, and then they use their limbs to propel themselves forward in the water. When the animals utilize their power stroke to propel their bodies forward, it is called a power stroke. To begin the stroke, animals use little water movement to propel their limbs, and this causes the animal to move a small amount backward. The resisting force on the animal as it returns to the water is known as drag.
- **Jet Propulsors**: Compressed fluid creates a cavity inside the body, and the cavity is filled by blowing the fluid out. This ejects the fluid, which moves the animal in the opposite direction. Each animal has a different propulsion system in jet propulsion mode. This means that in jellyfish, water is drawn from the back and then ejected via the rear. The body of scallops is composed of two distinct compartments. Fluid enters the front section of the scallop, then exits the back.

- **Undulatory Propulsors**: Robotic undulatory propellers, which imitate aquatic creatures' stretched and asymmetrical fins, are being developed to improve movement and maneuverability. To accommodate realistic technical constraints, most robotic undulatory propulsors limit the number of longfin rays that connect flexible fin surfaces as backbones and divide the total number of membrane-like segments among them in their propulsion systems.

Integrating Blockchain and IoT with drones can estimate the amount of propulsion required to operate the drone. Additionally, it keeps track of the drone's propulsion system with a timestamp and helps to compare the statistics which helps in improving the performance. Further, these statistics will be useful to predict the future environmental conditions for drone operation and ensure the operating requirements.

### 1.1. Blockchain and UUDs's Functioning Mode

Blockchain can ensure high performance at a lesser cost [133][134]. This section explains how Blockchain can ensure drone's functioning in their modes of operations. Details are presented as follows.

- **Diving Mode**: In driving mode, the power system is used to aid in the needs of the propulsion system. The engine may use three different levels of power in a "full" or "high" condition (FULL), or in a "medium" or "normal" state (MEDIUM), and three alternative power modes (NORMAL, REDUCED NORMAL, EMERGENCY) [63]. Blockchain can help in automating these power modes based on the drone's functionality. Blockchain can ensure (i) minimal required support facilities, (ii) high speed in operation and functionalities, (iii) long-range coverage with less power consumption, (iv) ensure lower system cost, and (iv) greater operational flexibility. Presently, the unmanned drone's functioning modes face difficulties as (i) limited communications, (ii) limited work capability, (iii) energy limitations, and (iv) lack of adaptive intelligence. With the integration of Blockchain technology, these properties can be ensured.
- **Rising Mode**: This mode entails continuous acceleration and keeps the drone at a constant depth until it reaches a certain minimum depth.
- **Floating Mode**: This mode uses the GPS to collect location data and transmits it to the Iridium system. When you're in wait mode, you should wait for any support boat to arrive, and you should have Wi-Fi on so that you can download new commands or software updates.
- **Emergency Mode**: This sub-mode indicates a severe propulsion system deterioration, such as when the heat source system or the turbine system is out of service. This mode would typically necessitate using the backup power system to fulfill power needs. A Blockchain-based system keeps a record of data from sensors attached to different units and helps create a bridge-Blockchain between other functioning units to operate smoothly. For example, the distinctive characteristics of the Oxley compartment lights in underwater drones include rating, battery back-up, and explosion-resistant casings. These features can be stored in a distributed network with data available over multiple places using Blockchain technology. Additionally, In addition to the optical elements intended to assist decrease tiredness, the maritime lighting may also be utilized strategically to enable the crew to maintain their daily sleep cycles. To prevent the devices from high ambient heat, they feature a sophisticated heat management system built-in. With the heat dynamics accounted for, the LEDs are protected, and therefore perform at their maximum. The properties of heat dynamics, heat management systems, and other subsystems can be stored in a network to observe the functionality constantly and in real-time. In this manner, Blockchain technology can help to automate operations.

Blockchain networks are more trustworthy compared to other security approaches. UUDs need sensors and IoT. Conventional IIoT designs say that because of a single point of failure and the fact that drones are generally simple targets for cyber-attacks, drones are at risk. The dependable and secure operations are assured by a Blockchain-based architecture. Sensor and actuator data access will be governed by a private and lightweight Blockchain infrastructure. Using a low-power processor (e.g. ARM Cortex-M4 CPU), the Blockchain network can perform real-time cryptographic computations using a highly scalable, quick, and energy-efficient consensus method. As a consequence, lightweight Blockchain architectures may be beneficial for drones. Figure 8 shows how lightweight Blockchain infrastructure can be integrated with a drone's mode of operation using bridge Blockchains. This way, Blockchain helps in switching between different modes while the drone is in operation.

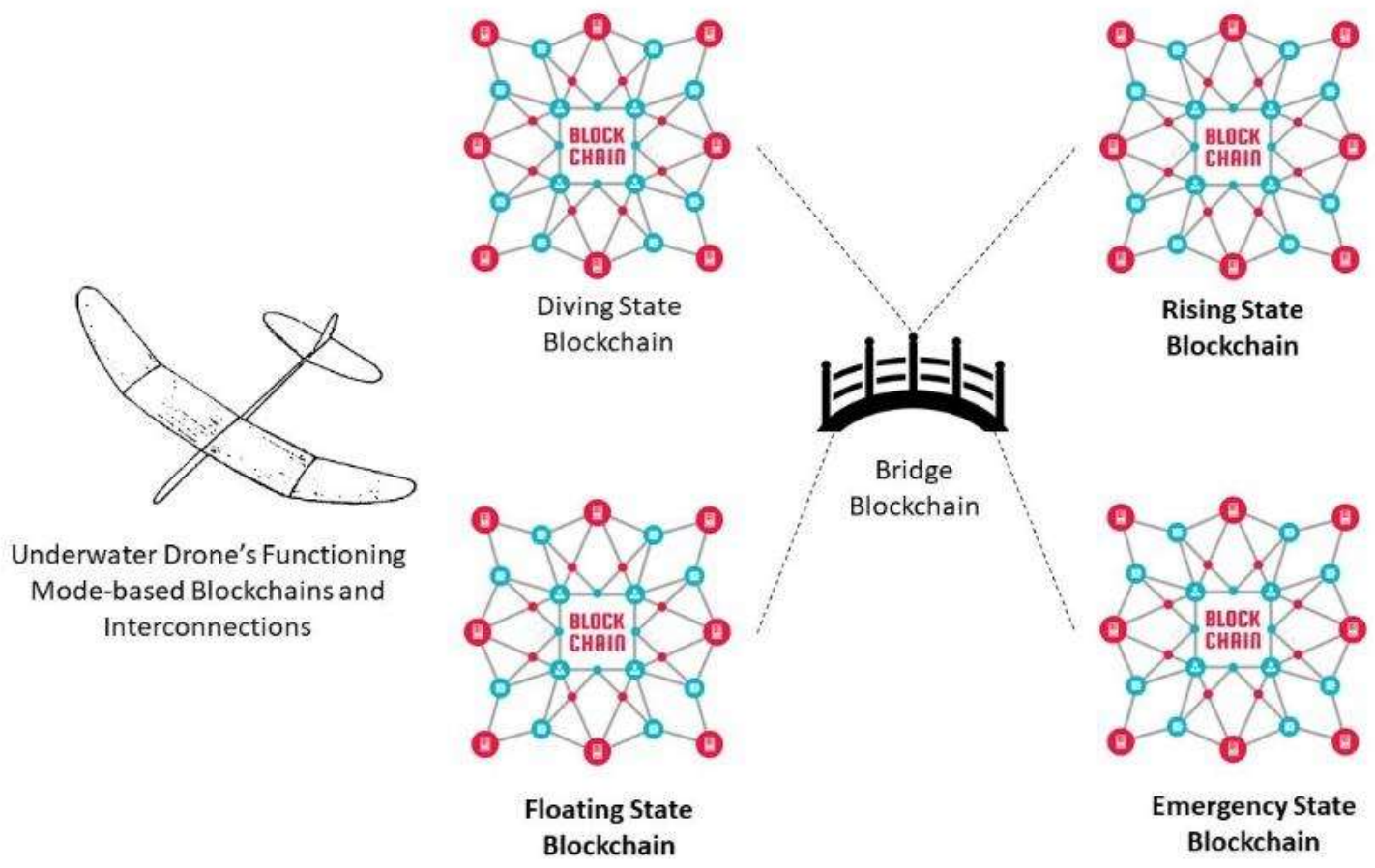

**Figure 8:** Underwater Drone's Functioning Mode-based Blockchain

## 6. BLOCKCHAIN USE CASES IN UUDs

Blockchain use cases in UUDs are the emerging applications to improve the research data and to provide the new data from very difficult conditions where humans couldn't sample data. The onboard sensors of underwater drones acquire data and the data is sent in real time via different modes. The use cases of Blockchain based UUDs are discussed under this section:

### 6.1. Blockchain-based Military Applications of UUDs

The military of the nation considers the security of the beach borders as equally essential as the security of the land borders. Massey et al. [64] identified that the borders can be monitored using aerial, water surface, underwater or ground surveillance, drone technologies. The security of the data captured from the UAVs or UUV is important as it is data related to the border security and is highly confidential. Dutta et al. [65] established a methodology of the immutable Blockchain ledger that provides the reliability of data without third parties involvement. The problem in a conventional system which relies on intermediary parties can be solved using a decentralized Blockchain platform. The main requirement for border security is to detect terrorism and intrusions using cost effective applications.

#### 6.1.1. Blockchain Architecture

In the previous studies the Blockchain ledger has not been deployed in underwater drones for border security and intrusion detection. A project ROBORDER is a border surveillance system developed with an objective to demonstrate a fully functional autonomous system using robotic swarms systems including aerial, water surface, underwater or ground surveillance drone technologies. Orfanidis et al. [66] discussed that Computer vision enhancement and enrichment of information semantically could be incorporated into the swarm intelligent system for the border monitoring purposes. However, the industrial applications of swarm robotics for border security have not been deployed successfully. Milanie et al. [67] provided information on various industrial projects where swarm robotics has been already implemented but the distributed decision making is completely neglected. The architecture for Blockchain based border intrusion detection is given as Figure 9.

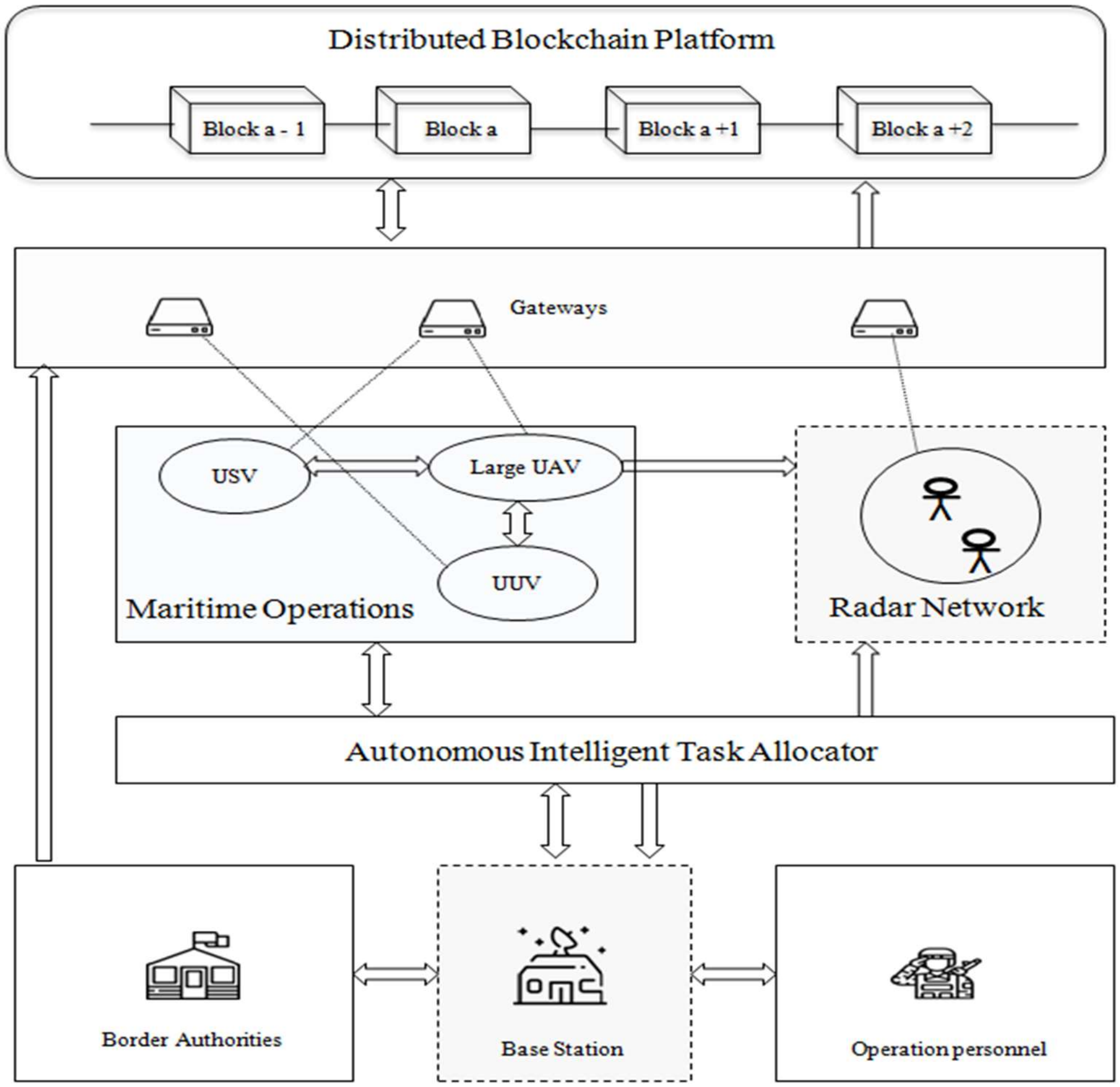

Figure 9: Blockchain based platform for Beach-border Identification

The autonomous task allocator based on Artificial Intelligent algorithms intelligently monitors the USV, UAV and UUDs surveillance, underwater movements and allocates tasks to be performed. These tasks are also informed to the authorities and operation personnel via base station. The data sensed through the underwater devices is updated with the radar information and to the Blockchain ledger through gateways.

#### 6.1.2. Challenges

The threats in the remote locations and patrolling the large area are the challenges related to the patrolling and protection of the borders. The major issues that are involved in the implementation of blockchain use cases in military services are discussed here:

- Security issues- the DoS attack can slow down the speed of processing of new blocks in the blockchain ledger and time can be spoofed to delay the communications. Moreover, frequent system updates require the integration of more expensive systems into the whole application.
- Scalability- the blockchain implemented into the system is not capable to adapt itself according to the small or large operations.
- Interoperability- the blockchains inability to share or operate with one another is an interoperability issue. For data sharing among blockchains to have similar identities between the two blockchains is required, and this feature is not present in blockchain technologies.
- Long term management and budget planning for blockchain implementation in UUDs is another issue. The government funding to deploy blockchain based UUDs would be expensive and risk taking at this initial stage of blockchain implementation for military services.

### 6.2. Blockchain-based Ocean Life and Monitoring-based Applications of UUDs

Blockchain based ocean life monitoring is a modern world application that can effectively solve multiple environmental issues. The rise in water level pollution and its effect on ocean life increases the serious environmental risks. The WHO presented the issues that affect the quality of water. These issues include infection with the pathogens Soprani et al. [68], contamination with organics Cai et al. [69], Acidification Shi et al. [70], Eutrophication Leaf et al. [71], Agricultural fertilizers that goes into the water bodies Bouraoui et al. [72], and heavy metal pollution Akpor et al [73]. The global water quality is at increased risk of deterioration and requires environmental monitoring. The system of monitoring of ocean life and water quality has to be transparent, in which the data monitored would be verified and is available to every citizen for their validation. The advantage of IOT sensors to collect and verify the gathered data makes the process more reliable and transparent. Berman et al. [74] and De limpa et al. [75] discussed about the sensors attached to the system that are capable to measure pH, DO percentage, ORP (mV), Conductivity (μS/cm), Temperature ($^{0}$C), Turbidity. The underwater drones that are built based on Blockchain along with highly effective IOT sensors are capable of providing high efficiency and reliability.

### 6.2.1. Blockchain Architecture

A Blockchain based distributed framework presented in Figure 10 is utilized to create a model to monitor, collect and update information in a secure manner for water quality assurance and other research-based monitoring purposes. Berman et al. [74] proposed an architecture for monitoring ocean life and water quality utilizing ethereum based UUDs.

#### 6.2.1.1. Utilized Technologies-

The technologies utilized in the implementation of the architecture to monitor ocean life and water quality are: The ROS, The IPFS, Liability Market, Liability Contracts, and Tokens. These technologies are briefly explained as follows.

- **The ROS**: Koubaa [76] discussed the interoperability of heterogeneous system architecture such as autonomous robotic systems and IOT system hardware and software interfaces to work in coordination, ROS is deployed into the system architecture.
- **The IPFS**: The IPFS file system allows working with distributed file systems for data storing and sharing purposes.
- **Liability Market**: Liability market transactions are controlled through the IPFS system. It has the responsibility to match the demands among the nodes in the system architecture.
- **Liability Contracts**: The Ethereum smart contracts with humans or within systems (AI, IoT and intelligent systems).
- **Tokens**: The **exchange** of currencies within a network is performed using a token mechanism.

The working of this architecture is a three-phase process. In the initial phase under the IPFS the promise node sends a demand message in the liability market. This demand is transferred from the provider to the agent to perform some tasks. CPS can accept an offer or submit a counter demand. This phase terminates when offer/demand messages are equal in all the fields. The next phase is the generation of a new contract in the Ethereum Blockchain. In the execution stage the AIRA software, when receiving the liability creation message, passes the acknowledgement to the agent. The CPS subscribes to the acknowledged ROS topics to get the required information, and then the execution phase begins. The third phase is the finalization stage, in which the CPS acknowledges the AIRA software of the task completion. AIRA software gathers all the operation logs information into the result message and sends it back to the IPFS system. A node validates the received result message. The provider at the end is notified by the result message and registers the transaction to the Ethereum.

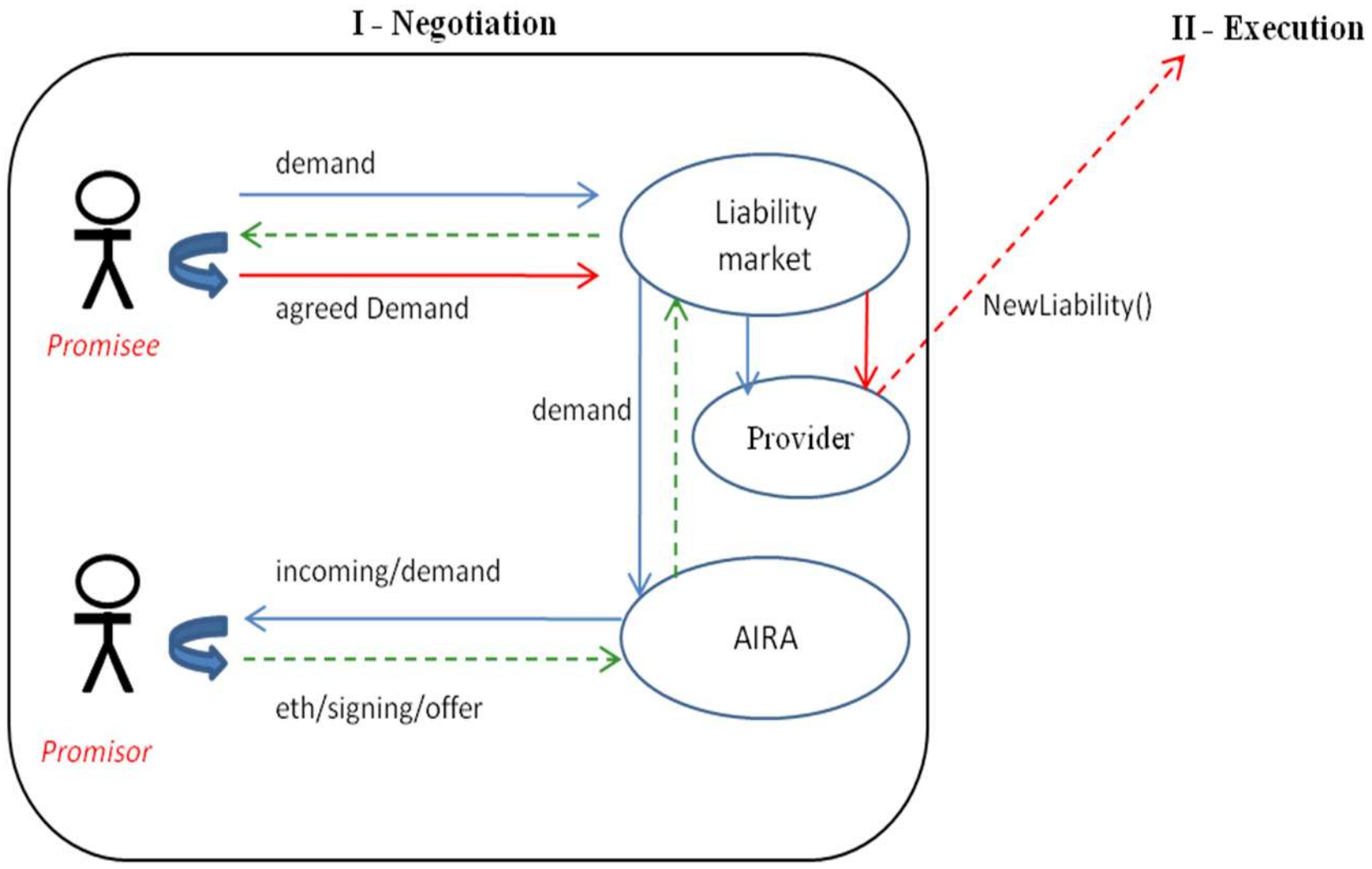

(i) Nodes interaction in Negotiation phase

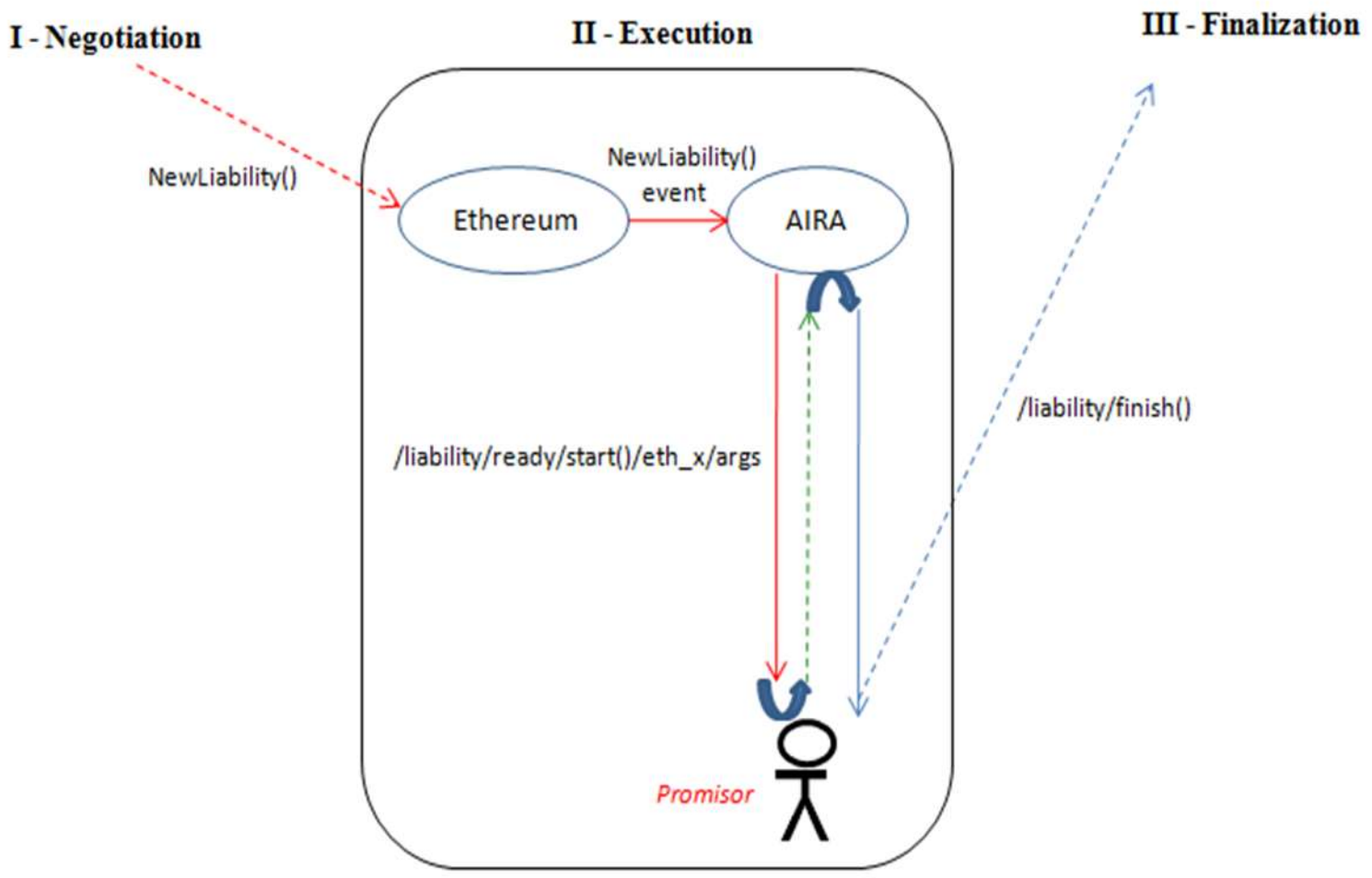

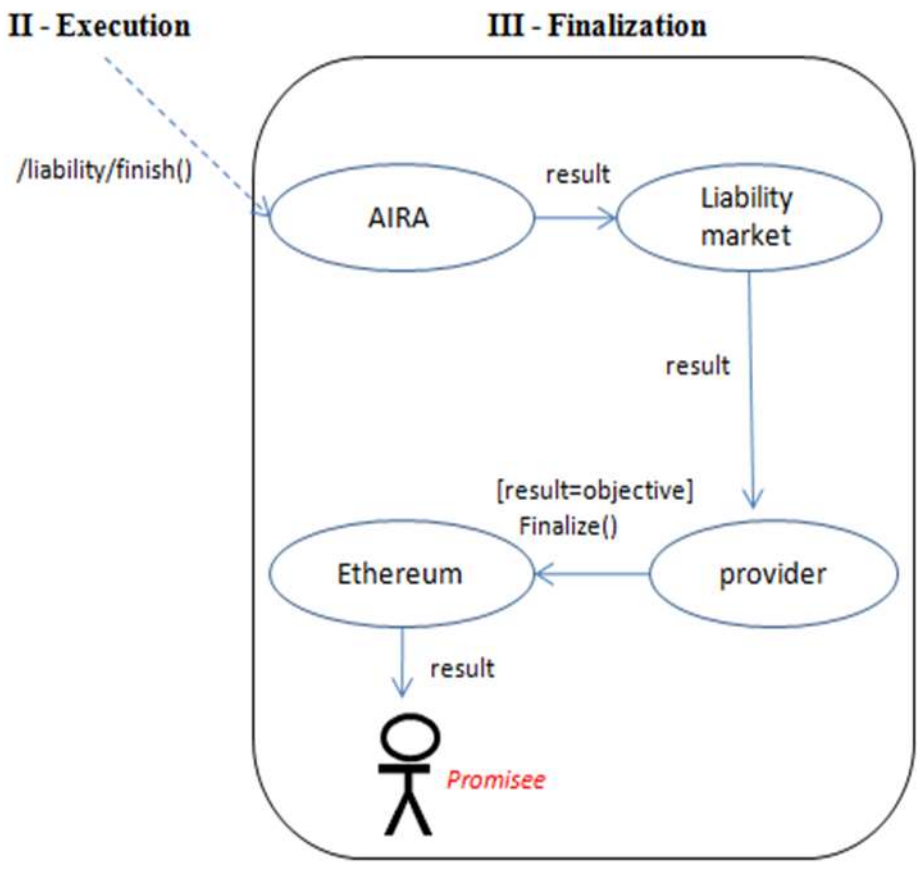

(ii) Nodes interaction in Execution phase in ethereum based monitoring system

(iii) Node interaction in Finalization phase

Figure 10: Nodes interaction (i) Negotiation, (ii) Execution and (iii) Finalization phases in Robonomics environment of offer and demand [74]

### 6.2.2. Challenges

The main threats related to ocean water and wildlife monitoring are Unskilled personnel,Security risks, and Energy consumption. the lack in technical knowledge limits the use of blockchain

- Unskilled personnel-In developing countries there is a lack of skilled personnel, this limits the growth of globally coordinated observing systems POGO.
- Security risks- the security risks increase when data is shared among multiple peers with limited cryptographic algorithmic protection.
- Energy consumption risks- blockchains consume enormous amounts of energy because of the algorithms used for its implementation.The increase in energy consumption increases the complexities in monitoring of large areas effectively.

## 6.3. Blockchain-based Ocean Wild Species Classification Approaches using UUDs

The monitoring of wild species in Oceans mitigates the illegal transaction and exploitation of ocean life. This process to save ocean wild life improves if these species are well classified. The classification of ocean wild species is generally based on their behavior, habitats, or food demands. This classification improves the data of the wildlife in the oceans, which indirectly helps research studies to benefit these species. Tawalbeh et al. [77] provided information on the lack of security of data gathered from IoT sensors, due to the involvement of mediators as a major drawback in the existing classification. Uddin et al. [29] presented a Blockchain based unmanned underwater vehicle that provides a distributed platform to save the data, which cannot be altered or manipulated by any other parties. In one study author Skomal et al. [78] classified the behavior of the whale as:

- *Travel*: the whale's group with consistent one direction movement, speed can vary (fast or slow) is classified under the class *Travel*.
- *Forage*: the other group of whales with no consistent directional movement, but rather confined movements to a common area are classified under the class *Forage*.
- *Social*: the whale's group that has demonstrated coordinate group activity are kept under the class *Social*. This group has been found to have a particular pattern in their daily activities.
- *Rest*: the whale's group where whales remain in the same place classified under the class *Rest*. This group of whales has performed overall less activities as compared to the other classified groups.
- *Unknown*: the other whale's group where no recognizable patterns are observed are considered as *Unknown*.

### 6.3.1. Blockchain Architecture

The incorporation of Blockchain into the classification process of wildlife species enhances the security of data acquired. The immutable data in blockchain makes the classification process of the ocean research data more reliable. The Blockchain architecture for wildlife classification process utilizing blockchain installed UUDs is depicted in Figure 11.

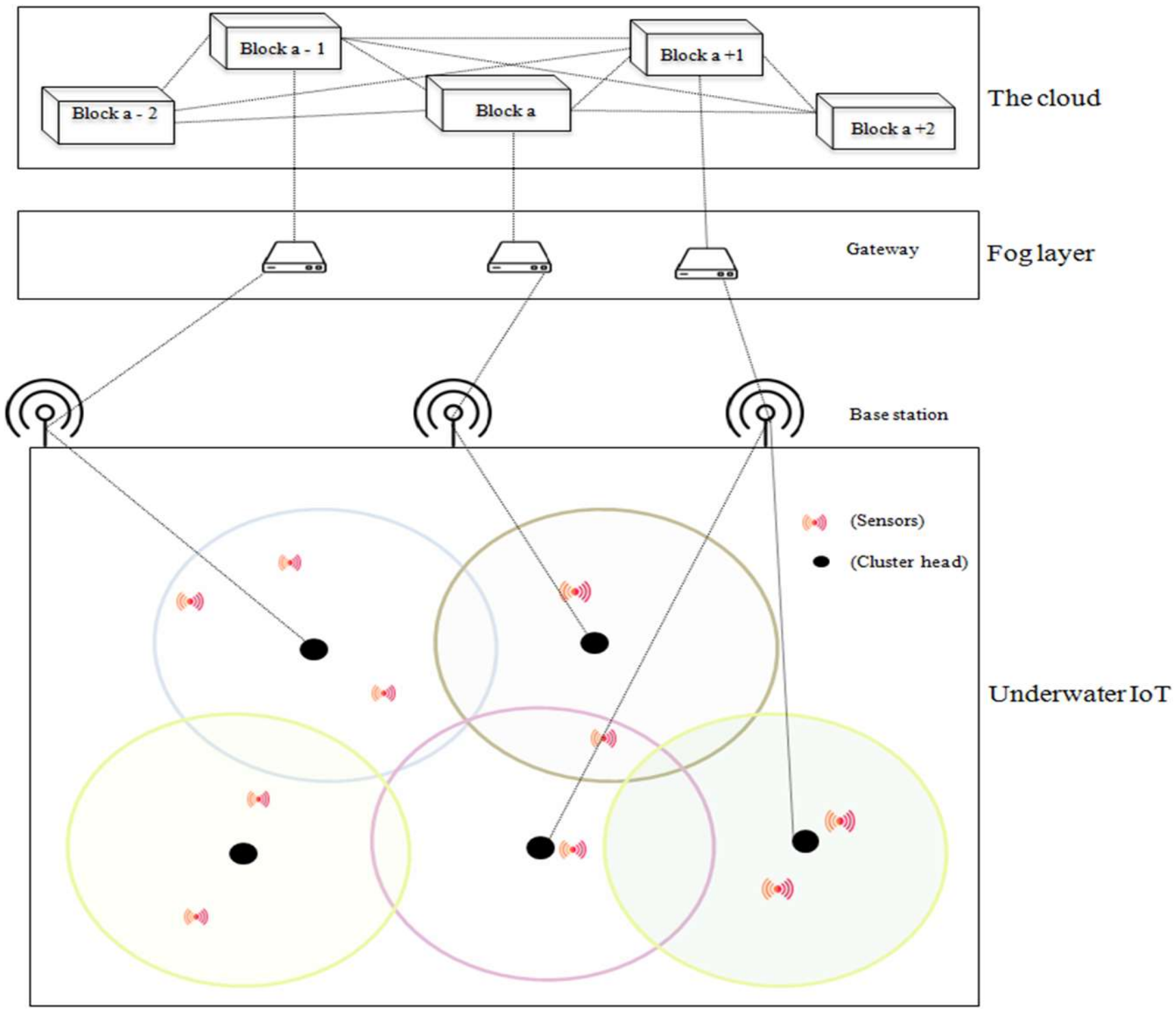

Figure 11: Blockchain based distributed platform for wild species classification

The underwater IoT layer in Figure 11 is deployed with underwater drones that are equipped with highly efficient IoT sensors and high-resolution cameras. The cluster is the subdivision of a large area to be observed and each cluster has some sensors to trace the movements, habitats and other significant details related to ocean research. These sensors with underwater drones are incorporated to the specific zones or clusters to navigate and collect the data related to the specific cluster. Each cluster comprises some sensors and UUVs, the cluster head is the prime sensor to collect all the information from the rest of the cluster sensors. The cluster head transfers the signals to the base station to connect to the distributed Blockchain ledger through gateway permission. The fog layer gateways connect the base station with the cloud layer Blockchain. The data gathered through sensors are then transferred to the blocks, data is saved as blocks which are immutable and secure with cryptographic algorithms.

### 6.3.2. Challenges

Various challenges in this area are briefly explained as follows.

- Skilled manpower- The use case illustrates that implementation of blockchain requires a high level of skilled and technological experts to handle the data on implemented blockchain efficiently.
- Adoption issue- the blockchain use case for classification purpose requires work to be implemented at a broad range with public trust. The public trust and support to adapt with the changing technologies is the main challenge.
- Data security- the wildlife classification algorithms contains sensitive information about the wild species. Although the blockchain is a secure technology, the classification algorithms used in the process are vulnerable to several attacks.

## 6.4. Blockchain-based Beach-border Identification approaches using underwater images and UUDs

The deployment of UUVs into the ocean with well-equipped infrastructure and blockchain technology is capable of serving multiple tasks: Monitoring, classification as well as for beach border identification approaches. The integration of advanced communication technology and underwater drones can support search and rescue operations at the beach-border. The UAVs and UUVs are deployed in the different water regions to collect and store the data to send it to ground control stations. The data is usually in the form of images and videos that are analyzed and observed at the base station with advanced algorithms. The algorithms are generally focused to identify humans, wild species, other UUVs and to detect other illegal activities at the beach-border. Papakonstantinou et al. [79] proposed computer vision algorithms for 3-D representation with image processing for data analysis. The fuzzy classification used statistical features as input to classify output classes of coastal composition as (sand, rubble, rocks) and sub-surface classes as (seagrass, sand, algae, rocks). Although the blockchain has not been incorporated in this study, the output of the work is quite significant. The implementation of a blockchain based data collection process can provide a reliable, transparent and secure system and could also improve the overall effectiveness.

### 6.4.1. Blockchain Architecture

The blockchain architecture of beach-border Identification using underwater images is depicted in Figure 12. The beach borders categorization according to its climate, beach composition and subsurface environment is a complex process. It requires deployment of UUVs, UAVs equipped with high end resolution, fully stabilized cameras capable of performing effectively also in the extreme conditions. The quality of images captured determines the data quality and overall decision-making process. Schettini et al. [80] discussed the methods to improve the range of underwater imaging, improve image contrast and resolution. Additionally, the images gathered are highly confidential and require a reliable, immutable, transparent system for storing the image data. The blockchain based technology, if combined with the pre-developed data gathering systems, will improve the data quality significantly.

The blockchain based distributed platform for beach border identification approaches using underwater images collected through underwater drones is depicted through Figure 12. The architecture includes the UUVs equipped with high quality, and fully stabilized cameras deployed at different regions of the beach border. The data is captured and sent to the base stations and from the base station to the distributed blockchain ledger through gateways.

### 6.4.2. Challenges

The main challenges associated with beach border identification utilizing underwater images and videos are energy issues, financial constraints, user's trust and interoperability.

- Energy issues- the energy consumption of the overall ethereum network uses as much energy as an average household per week.

- Financial constraints- Blockchain implementation for many organizations is not feasible due to the requirement of high budgets.
- User's trust- lack of trust among blockchain users has restricted the blockchain deployments in industrial sectors and other organizations.
- Interoperability- there are no standards formulated that enable different blockchains to communicate with each other in an organization.

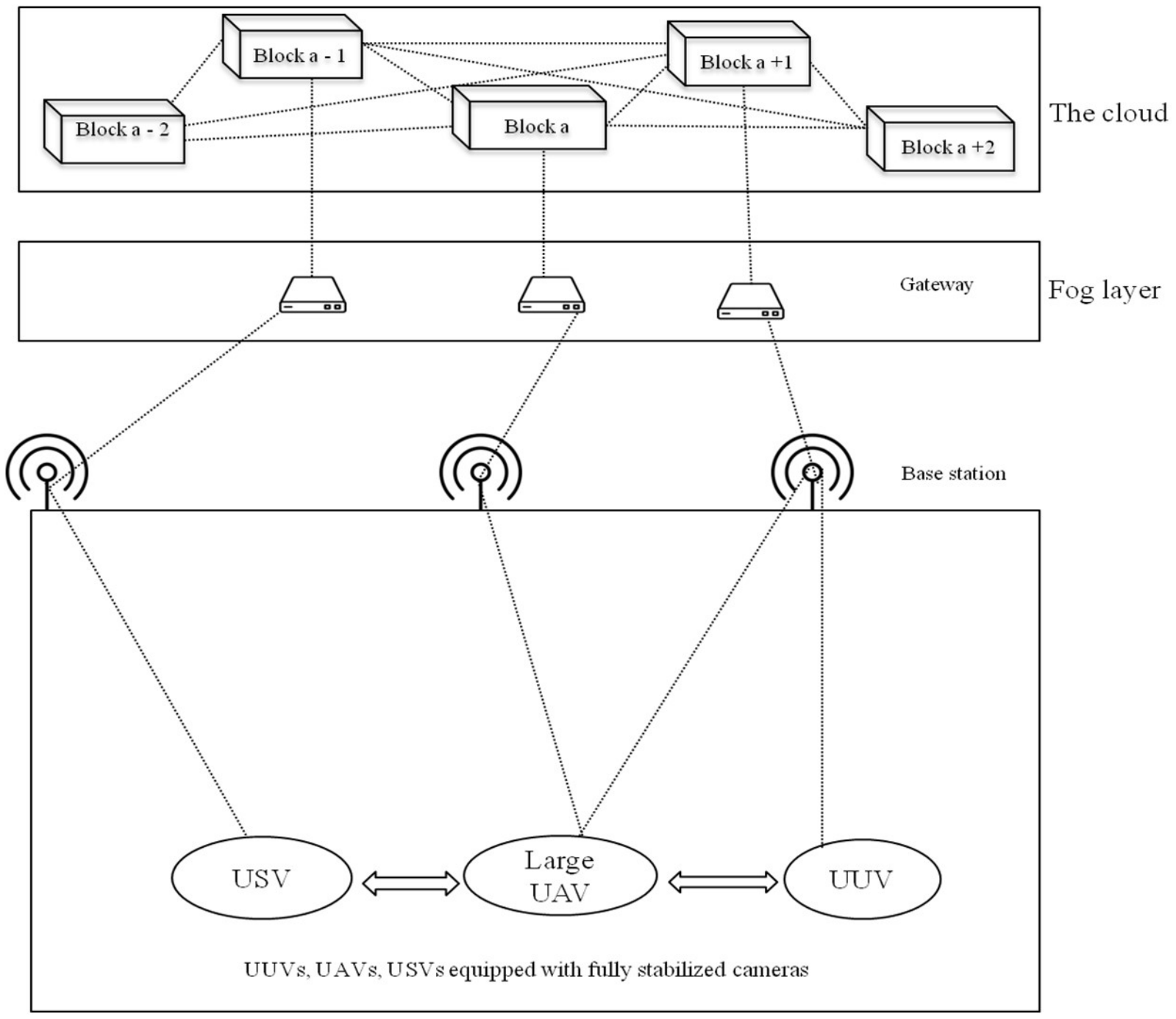

**Figure 12:** Blockchain-based distributed platform for beach border identification approaches

There are advancements in the field of research based on underwater drones and blockchain technologies. Although underwater drones and blockchain are highly capable technologies, there are only a few studies that have incorporated the benefits of both technologies together. The blockchain architecture in some case studies is only to present an idea of integrating two technologies effectively. Some projects and research based on underwater drone technologies are presented in Table 3.

Table 3: Projects and research classification of underwater drones

| **Survey** | **Number of robots** | **Environment** | **Drone type** | **Behavior** | **Blockchain -based Platform (Yes/No)** |
|---|---|---|---|---|---|
| **Military** | | | | | |
| OFFSET | 250 | Aerial | UGV, UAV | Navigation, Decision making, Miscellaneous | No, the architecture is not based on blockchain algorithms. There is risk on the use of countermeasures like electronic warfare techniques, cyber-attacks, laser and microwave weapon systems |
| Perdix [81] | 103 | Aerial | UAV | Navigation, Decision Making, Spatial organization,Miscellaneous | No, The architecture is not based on blockchain algorithms. Moreover, the drone swarm would require mission specific programming. |
| CARACaS [82] | 5 | Aquatic | USV | Navigation, Spatial organization | No, the interoperable system is centralized and is not based on blockchain algorithms. |
| Hercules [83] | N.A | Aquatic | UUD | Navigation, Miscellaneous | Yes, the architecture is based on blockchain algorithms and the network is distributed. There is no risk on the data security. |
| **Environmental Monitoring** | | | | | |
| CoCoRo[84] | 41 | Aquatic | UUV | Navigation, Decision making, spatial organization | No, the architecture is not based on blockchain algorithms. |
| CORATAM [85] | 12 | Aquatic | USV | Navigation, spatial organization | No, the architecture is not based on blockchain algorithms. The data is on the risk of cyber attacks. |
| Platypus [86] | 25 | Aquatic | USV | Navigation, Miscellaneous | No, the architecture is not based on blockchain algorithms. |
| Apium Data Diver [87] | 50 | Aquatic | USV, UUV | Navigation, Spatial organization, Miscellaneous | No, the architecture is not based on blockchain algorithms. |
| subCULTron [88] | N.A | Aquatic | UUV | Navigation, Spatial Organization | No, the architecture is not based on blockchain algorithms. The programming of the system is mission specific. |
| Vertex Swarm [89] | 10 | Aquatic | UUV | Navigation, Spatial organization | No, the architecture is not based on blockchain algorithms. |
| SWARMs [90] | 8 | Aquatic | UUV, USV | Navigation, Spatial organization | No, the programming is mission specific and is not based on blockchain for data security. |
| **Surveillance** | | | | | |
| ROBORDER [91] | N.A | Terrestrial/Aerial/Aquatic | UxV (UGV, UAV, USV) | N.A | No, the architecture is centralized and is not based on blockchain algorithms. |
| GEOMAR-ABYSS, ACFR-SIRIUS [92] | N.A | Aquatic | UUD | N.A | Yes, the architecture is based on blockchain algorithms. There is no data insecurity in distributed platforms like blockchain. |

## 7. INTEGRATION OF BLOCKCHAIN AND ENABLING TECHNOLOGIES FOR UUDs

UUDs are more challenging and complex than Unmanned Aerial Drones. Unlike air, a water surface can behave like a mirror-surface, diffused surface or transparent surface. This behaviour creates a specular reflection which leads to a technical challenge to ML algorithms. Carrivick et al. [93] overcame these problems in visibility and distortions in the images received underwater by reprocessing of data using SFM workflow.

Butcher et al. [24] emphasized upon the main factors affecting the data collection in this UUD like problems to access, geolocation, weather, wind, waves and surface visibility. Besides, Depth, Patrol time and Turbidity also studied which affect the drones' motion underwater. The role of AI in this complex environment is beneficial in solving the problem of speed, accuracy of collected data and data security. This data can be in the form of image, video, sound and real-time interaction. Implementation of Blockchain and ML algorithms overcome these technical issues.

The study of Blockchain algorithms and ML algorithms for UUD can be categorised among the following types based on network sharing:

- Wireless Underwater Drone / Robotic/autonomous Underwater Drone/ Bluetooth based Underwater Drones – These robotic drones are pre-programmed and need neither human to operate nor real-time input. These devices can be propelled or non-propelled.
- Tethered Underwater Drones / RODr – These devices are connected with wires and are manually operated by the operator.

UUD store data in various forms which is collected and processed. Multiple drones create a swarm and can collectively perform as a single unit. Combination of AI, cloud computing, Blockchain improves the efficiency of these drones. Wireless underwater drones are better compared to ROD's in terms of driving distance and time interval [94].

### 7.1. Blockchain and AI/ML for UUDs

Blockchain works as a decentralised network consisting of blocks with a record of transactions/data connecting with a previous and successive block having a hash code that restricts any alteration in the data without alteration of all connected blocks. There is no centralised authority and the authenticity of data is through a collective form of blocks. The use of ML makes the data collection cost-efficient and ready to use for scientific purposes in real-time [24]. A Consensus Algorithm in Blockchain deals with Data validity, data privacy, data verification, data security and data transparency. All blocks collectively verify the records. It ensures that each new block has an agreement with all previous blocks.

Validation-based Blockchain consensus algorithms are:

1. PoW: It produces a cryptogenic hash for data validation. This is a permissionless Blockchain type in which mining is based on computation power. Each time the input data runs, it adds up an arbitrary number. High power is required to add data as a block in the Blockchain. Specialised computers solve complex problems required by systems working in PoW. Bitcoin introduced by Satoshu Nakamato works on this algorithm. [95],[96],[97]
2. PBFT: This is a Private Blockchain type developed to work efficiently in asynchronous systems. This has low energy consumption. It can work with malicious nodes if they are less than one third. [97][98]
3. PoS: This is a public Blockchain type similar to PoW except for the validation process. It reduces the electricity consumption as compared to PoW. Here, based on participants' stake on networks these blocks are validated as tokens. This is a permissionless Blockchain type in which mining is based on validation. World's largest Blockchain network Ethereum uses this algorithm[96][97].
4. PoET: This is developed by Intel. It relies on a timer system. To create a new block, network participants should complete the given time-frame. After validation, the participant gains a new block. PoET is used by private blockchain [97].

Voting-based Blockchain consensus algorithms are:

1. PoV: This is a consortium-based Blockchain type with vote-based mining. It has a decentralised form for book keeping and voting. The consortium members vote for the validation of the consensus node. With majority votes, i.e., more than 50% votes, validation and the verification of nodes is achieved. The PoV has butler and commissioner nodes. Commissioner nodes ensures the security and trust environment [96][99]
2. Proof of Trust: This is a permission-based Blockchain type with Probability and Vote based mining. To create a trustful environment, the peer role is included as validating nodes. This peer network is valued in this decentralised system. This reduces the work and saves energy consumption among peers [96].

Authentication based Blockchain consensus algorithm:

1. PoA: PoA is based on trustworthy parties and network participants. It validates nodes chosen by these parties. The existing parties have fixed blocks and they validate the transactions. They behave as validating nodes. It can be utilised in supply chains and trade networks [97].

2. PoP: Mohanty et al [95] introduce a new blockchain architecture for data management and data security by integrating hardware, blockchain, IoTs and data security. They created a Physical Unclonable function PUFchain. PoP is 1000 times faster than PoW and is the integration of PuF and Proof of Authentication.

Blockchain has a distributed database that increases testing and helps in tracking and monitoring effectively [100]. There is an increase in the research and development of UUD in various areas like Military applications, Ocean life monitoring (Flora and Fauna), Security, Beach-Border Identification, Surveys, Spying and Weapons. There are some research challenges, though. Some common research challenges for UUD's are High Pressure undersea, target accessibility, Diving hours, absence of GPS, Absence of light at a deep level and high amount of light scattering. These challenges are overcome by implementing concepts of AI, ML Algorithms and Blockchain. Image processing has a huge research contribution in UUD studies to make it work in real-time. Table 4 below shows the Types of UUD's and their characteristics:

**Table 4:** Types of UUDs and their characteristics

| **Types of UUD** | **Sub-Types of UUD** | **Examples** | **Architecture** | **UUD Technique Features** | **ML algorithms** | **Blockchain Algorithms** | **Applications** |
|---|---|---|---|---|---|---|---|
| Wireless UUD | Bluetooth based UUD, Autonomous/ Robotic UUD (Propelled and Non-Propelled) | GEOMAR-ABYSS, ACFR-SIRIUS [24] [92] | Hierarchical, Heterarchical, Subsumption, and Hybrid architecture | Video, cameras, Sonar, Magnetometer, fluorometer, oxygen sensor, conductivity, temperature, depth sensor, pH sensors | • KNN<br>• Decision Tree<br>• SVM<br>• Hidden Markov Model<br>• Random Forest<br>• Regression<br>• Naïve Bayes<br>• CNN<br>• ANN<br>• Deep learning | • PoS<br>• PoA<br>• PBFT<br>• PoV<br>• PoT<br>• PoET<br>• PoW | • Oil and Gas location exploring<br>• Military application<br>• Ocean life monitoring<br>• Beach Border Identification<br>• Spying<br>• Weapons<br>• Security<br>• Lake/Dam inspections<br>• Pipelines inspections |
| Tethered UUD | Remotely Operated Drones | Hyper Dolphin [52], Hercules [101]<br><br>iBubble | Swarms, Control, Survey, Video, Multibeam | Monitoring, Data collection, Thermal Image, Object Detection | | | |

Some most implemented ML algorithm are K-nearest neighbour, SVM, Hidden Markov Model, Decision Tree, Random Forest, Regression, Naïve Bayes and Graph Theory. Some major contributions are following:

- Carrivick et al. [93] uses SFM data to quantify underwater structures and aquatic life with the help of drone technology. Computer vision and photogrammetry are used to produce 3D clouds. Real life implications of Sfm with MVS using multiple sensors were studied for the exploration of the unexplored part of marine life. For future research in fluvial hydraulics an improved quality of reconstructed water surface would be a mile stone. In addition,it is now possible to use the satellite images, hydraulics and river bed evolutions along with SfM and MVS and measure the image data more accurately and highly detailed.
- Kwasnitschka et al. [25] emphasized on the factors affecting access, locations, visibility and drones' motion underwater. They proposed an improved optical system for visual mapping of sea floor while overcoming the limitations of drone technology like light scattering, high pressure and low visibility. Authors presented the system capable of operation at depths of 6000 m and construct 3 D high resolution images. Coloured High resolution underwater images were received using the created system. The system overcomes most of the underwater challenges of underwater drone technology.
- Chamola et al. [100] proposed integration of drone technology and Blockchain in various devices during Covid-19 situation helps in crowd surveillance, screening masses, broadcasting information, contact tracing and delivering medical supplies. DAG data structure along with deep learning was used in Blockchain based apps. Use of consensus algorithms minimise the data delicacy.

- Valavanis et al. [92] were curious about the problems with underwater vehicles in Ocean industry. They reviewed 25 existing AUVs and 11 control architecture system. The problems they found most affecting were consumptions of power sources, navigation and signal problems in underwater and management of these systems. After finding the problems with Compact PCI, they proposed architecture for a sensor based embedded underwater control system. They proposed QNX real time OS and single board compilers STD 32 SBC.
- Roman et al [101] in their attempt to study the underwater archaeological sites find the problems in creating high-resolution bathymetric maps. They compared laser profile imaging and stereo imaging. Acoustic and photographic mapping techniques are also used to document the texture and structures of underwater archaeological sites. In their work, remotely operated vehicle is used to generate bathymetric mapps of underwater archaeological site. Out of three different sensors; laser scanning, multibeam and stereo photogammetry, the first two were consistent.Meng et al. [102] provide an architecture for fish classification. They studied underwater drones and experimented with recognition of marine life in a natural lake. For implementation, the underwater drones captured images. To get a 360-degree panoramic view underwater, they used a fisheye lens. They succeeded in getting an 87% accuracy for recognizing species of fishes using CNN and Deep learning by a UUD enabled with Panoramic Camera. They achieve 360-degree view with getting real-time images from two 235-degree fish eye lens, Raspberry Pi compute model, Open SCAD as frame and deep learning were integrated.to get these results.
- Pham et al. [52] uses multiple drone setup and implemented consensus algorithms, image processing, avoidance obstacle algorithms and collision avoidance algorithms to study multiple coordination and formation control in real time. They used an open-source environment. Robotic model in gazebo simulation and in real world are studied. implemented DANCC to form a leader follower approach in multiple drone settings. Radial basis function and a potential function ensures that no collision should take place among the drones. A time variant function makes it simple to change position within drones without collision. The implementation is done on Gazebo underwater drone. Problem statement is developed using Graph Theory and Neural networks.
- Khelifi et al. [103] emphasized the importance of data security and secure content forwarding. Authors used integration of Blockchain in Vehicular Named Data network and Graph theory. They introduced a secure discovery protocol BC-VNDN based on PoW consensus algorithm. Intelligent attacks are possible in content-based security and in-networking systems. Data privacy and data security are also a concern and challenges. The authors proposed a reputation-based mechanism. The proposed model is tested in a simulation. It forwards and delivers only valid and trusted information.
- Mendiboure et al. [104] introduces the concept of the Internet of Vehicles. The vehicles can be connected to each other and surrounding devices from environments like pedestrian smartphones phones, road signals Data privacy, Security, a trustworthy environment, and a strong vehicular network needs to be explored. In future work, the authors suggest the concept of the Green blockchain. A Green block chain implies reduced energy consumption and improved storage abilities. Authors suggest the incorporation of AI, Software Defined Network and Edge computing in resolving the transparency and security issues
- Liawatimena et al. [105] uses deep learning and computer vision to improve the fishing management. Authors applied CNN and Computer vision to identify the fish species and population in Indonesia. The height and depth of the drones underwater, types of fishes available, their shape, size etc and monitoring the time of the day using Computer Vision helped in managing fishing resources. Authors with their results manages the fisheries and reduces the fuel cost on fishing trips. These UUD also helped in managing the traffic operations and estimating the tuna caught.
- Sparavigna et al. [106] discussed the use of Retinex filter and algorithm of GIMP to improve the image visibility in night or during fog. Retinex filter are able to improve shadow areas of image. GIMP retinex filter is capable of image manipulation and can be used underwater drones. The areas of Underwater engineering, underwater exploration, search and rescue, ocean studies and river bed exploration have to deal with low quality or blurred images. Retinex is able to simulate human vision and convert the captured image into a real-life scene. The underwater architecture or structures which requires inspection, monitoring and maintenance can be operated with underwater drones with Retinex algorithms in Image processing.

Figure 13 shows the representation of architecture for UUD.

### 7.2. Blockchain and Image Processing for UUDs

UUD are mini submarines working on the principle of buoyancy. Sending radio signals is a crucial task. Use of Line, long tether and Bluetooth helps in controlling the vehicle movement and the camera functions. Refraction and scattering affect image formation. Refraction, attenuation and scattering degrade the visibility and image formation. Kwasnitschka et al. [107] worked on the challenges faced by Underwater drones while studying oceans. High water pressures, accessibility, location, blurred images, distortion in images and energy consumption are dealt in their study. Authors present an optical system for Deep-ocean study able to go 6000m and creating colour high resolution images. The pressure in UUD was reduced using LED clusters. Study of a volcanic terrain is in future scope and long-term goals.

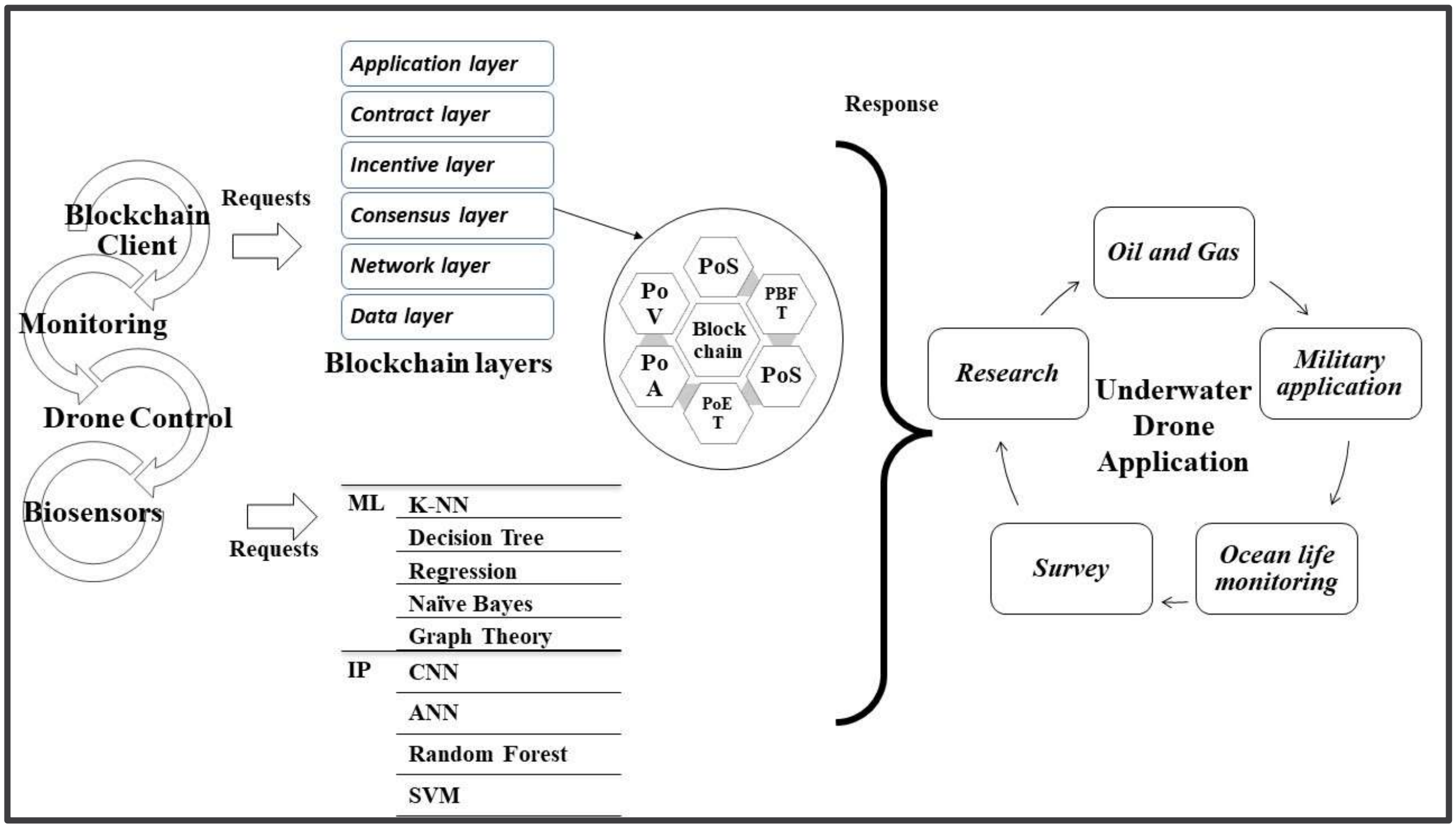

**Figure 13:** UUD Architecture

Higher functioning low-weighted cameras with features more compatible with the underwater environment are used in these drones to get higher resolutions.[24] Low-Cost underwater platforms and sensors are available with the ability to see and collect data and images in both fresh and salty water reservoirs [94]. Blockchain and Image processing helps to get a high-definition transmission of data, images and other signals in real-time. Thermal infrared, multispectral and hyperspectral systems are used in these UUD. UUD uses ML algorithms in Image Processing using ANN, CNN, Deep Learning and Computer Vision

Maire et al. [108] used CNN in marine life. They used it to detect dugongs using three convolution blocks. The use of computer vision in detecting larger data in the form of images along with Deep learning leads to better accuracy. The system was able to recall 80% of data correctly and precision improved to 27%. FishID, learning software, used an Object detection algorithm in tracking and identifying fish from data in video form. Hughes and Burghardt identify different species of sharks using Computer Vision and Random Forest along with Bayesian nearest-neighbour classifier. It gives a precision of 81%. Deep learning is used in fish recognition with the help of two fisheye lenses. 360-degree panoramic images were received with 87% accuracy. The drone was designed using Raspberry Pi, OpenSCAD and MakePro [28]. Clever Buoy is developed using a SONAR transducer to detect moving objects to identify sharks. Robotic Underwater vehicles can have various bio-sensors, including cameras, sonar, magnetometer, fluorometers, oxygen sensors, conductivity, temperature and depth sensors, pH sensors and turbidity sensors and GPS.

Major Research Challenges in Blockchain are lack of scalability, high latency, high energy consumption, fake block generation, stability, robustness and propulsion [95]. Table 5 below shows the ML and Blockchain Algorithms and their strength and weaknesses.

Table 5: Used Algorithms and their strength and limitations

| Algorithms | Algorithms used in Image Processing | References | Strength of used algorithm | Weaknesses of used algorithm |
|---|---|---|---|---|
| AI / ML | CNN<br>K-NN<br>Decision tree<br>Random Forest<br>Deep Learning<br>ANN<br>Computer Vision<br>Graph Theory | [100] [102] [52] [103][105][106][109][108] | Feature detection<br>Pre-programmed<br>No need of real time input | Need large amount of data to train |
| Blockchain<br>Consensus Algorithm | PoW<br>PBFT<br>PoS<br>PoV<br>Proof of Trust<br>PoET<br>PoA | [23][52] [94] [110] | Easy to verify the correctness of the solution<br>Secure<br>Decentralised<br>It can work with malicious nodes.<br>It is energy efficient.<br>Easy confirmations of transactions.<br>Low reward<br>Required low energy consumption | High Energy consumption. If a controlling group own 51% of data, they may corrupt the data.<br>It is used with combination of other algorithms to avoid the Sybil attack in which on party have multiple identities.<br>Participation is less. |

Sante Francesco Rende et al. [109] uses acoustic, visual and OBIA to build a seagrass habitat mapping in shallow and deep ocean water. They used Computer vision and different classification algorithms KNN, Random Forest and Decision tree to create thematic maps. Their integrated system is time and cost-effective. Multispectral images from various sources along with acoustics, videos and sonar system were studied to inspect the seafloor from coastline to deep water. Multiple data sensors yield better results with high accuracy and high-resolution images. It saves the monitoring cost-efficiency and improves the quality in comparison to single sensors.

Walter et al. [111] with underwater drone studied the link between tectonic fields and Geysir geothermal field using infrared mapping and Image Processing. The authors succeed to get datasets with temperature variations and fracture control for the Geyser geothermal field of Iceland through an underwater photogrammetry survey. The mapping through drones identifies the spatial distribution of hotspots, establish the link between these tectonic activities and dynamics of Geysers.

Lima et al. [75] studied the water quality and environmental impact under world largest solar panel park using UUDs with quality sensors. A water quality profile is generated by collective data. These large solar panel requires monitoring which depends upon the size, location and climate. In this study the impact was studied underwater and at open water. Sensors at different location and depth measure temperature, electrical conductivity, oxygen level and pressure every 15 minute. As future suggested works, authors emphasized that these studies need a large amount of time to figure out the impact on biodiversity. Intelligent UUD devices can be used for multi-sensory data collection and to plan certain underwater activities.

Alladi et al. [23] in their review paper studied about the blockchain in UAV's for military and commercial areas. Use of consensus Blockchain algorithms ensures collision free movements of multi-drones, uniform load sharing, data authenticity and fast synchronisations. Block chain offers the solution to the data security and communication problems in UAV. There is still lack of research in reducing the vulnerability towards security attacks and increasing the battery capacity and fly time. Battery life become crucial task as blockchain based system consumes more power

Detailed literature review of UUDs and their objectives, approach, technology used and limitations mapped with Framework given (A), Implementation done (B) and Blockchain focused (C) are given below in Table 6.

Table 6: Literature Review on recent UUD's

| **Authors** | **Year** | **Objectives** | **Approach** | **Technology Used** | **A** | **B** | **C** | **Limitation** |
|---|---|---|---|---|---|---|---|---|
| Gupta et al. [112] | 2021 | To reduce high data storage cost, network latency, reliability and bandwidth issues | Integration of Interplanetary file system and Blockchain | Blockchain based secure communication over 6 G network | Yes | Yes | Yes | Real time efficient deployment |
| Kozlova et al. [113] | 2021 | To design a UUD hull to study Arctic region | Mathematical Tools: Programming method, statistics, process simulation, Graph method, queuing theory | Life cycle management of a 3D printer Technology | Yes | **No** | **No** | Switch to additive technologies is not worth it. It is costly |
| Ramaswamy [114] | 2021 | To provide connectivity through drones during communication failure during disasters | Integration of Aerial Communication and Blockchain | Blockchain technology | Yes | **No** | **Yes** | Storage constraints, Energy Consumption, Poor network connection |
| Jan et al. [115] | 2021 | A theoretical framework for health care industry | IoMT | Blockchain and security | Yes | No | Yes | Higher maintenance cost |
| Trembanis et al. [116] | 2021 | Coastal Mapping and monitoring | Satellite imagery, Real time GPS, RADAR, Ground Penetrating RADAR, Light Detection and Ranging, SONAR, Acoustic Doppler Current Profiler | Blockchain, Deep Learning, Surface Motion Photogammetry | Yes | Yes | No | The efficient studies of coastal areas require a variety of sensors like Aerial vehicle, surface vehicle, underwater vehicle, UUD with geophysical sensor. |
| Lima et al. [108] | 2021 | Water quality observation in a solar park (before and after)and in open water | Water quality sensors measuring Temperature, oxygen meter, conductivity | Underwater drone for vertical profiling of water quality | Yes | Yes | No | Complexities in sufficient data collection in real time |
| Liawatimena et al. [105] | 2020 | To support fish management in Indonesia | Determining shape, size, colour of fish and analysis in Fish Aggregating Devices | Compute vision, CNN, Deep Learning, Cloud computing | Yes | Yes | No | Suited results are found with tethered drones. _ |
| Berman et al. [74] | 2020 | Environment Monitoring | Robot swarms based on Belief Space Planning, implementation of Reynold's Boids, Distributed ledger tools | Blockchain and Sensor networks on swarming Marine Vesssels | Yes | Yes | Yes | NA |
| Aggarwal et al. [117] | 2020 | Finding an optimal path | RRT fixed node, Reinforcement learning and E-Spral and E-BF | Djikstra's Algorithm, Q learning technique | Yes | No | No | Path planning given for UAV. Implementation on UUD is not suggested |
| Chowdhary et al. [118] | 2020 | To reduce the energy consumption of mobile Wireless network | Movement Score based Limited Grid-mobility approach | Reverse Glowworm Swarm Optimization Algorithm | Yes | Yes | No | search area taken is 2D space, collision free movement is considered |
| Papakonstantinou et al.[119] | 2020 | To produce marine habitat mapping | Comparing True Colour and Multispectral imagery | OBIA, k-NN, Fuzzy | Yes | Yes | No | Large differences in actual Multispectral images |
| Rende et al. [109] | 2020 | Seagrass habitat maping | Combintion of satellite multispectral image and underwater photogammetry data | Acoustic, optical data and OBIA, k-NN, Random Tree | Yes | Yes | No | Implementation over wide area is expensive and time consuming. |
| Mendiboure et al. [52] | 2020 | Formation tracking problem | Distributive Adaptive Neural Network Control | ROS, Gazebo, Graph Theory | Yes | Yes | No | Only 3 UUD were taken in study |
| Khelifi et al. [103] | 2020 | Blockchain system to secure Data forwarding and content caching | Named Data Networking | Blockchain and Vehicular Ad-Hoc Network | Yes | Yes | Yes | Non trust nodes can create privacy issue |
| Taber et al. [120] | 2019 | Mapping and monitoring coral reef | SFM and data volume | GPS-guided drone | Yes | Yes | No | Good Lighting system required for Image clearity |
| Meng et al. [102] | 2018 | Automatic fish recognition | fish eye lenses for 360 degree panoramic view | Deep learning, Raspberry Pi, OpenSCAD | Yes | Yes | No | This UUD has a speed of 2 knots and dive in 100 m. |

### 7.3. Blockchain and Future Generation Networks (5G, 6G) for UUDs

UUDs have provided cost-effective solutions to several societal problems, ranging from monitoring, surveillance, video-coverage, to health care, supply chain and many others. UUD have proved extremely useful during the covid 19 outbreak. The transportation of medicines, food supplies, and essential goods to designated locations within quarantined areas, undertaken by drones, turned out to be a valuable technology in combating covid-19 pandemic. Blockchain-based solutions provide privacy and security but suffer from issues of high storage cost, high bandwidth cost, low network performance and low reliability. Gupta et al. [112] proposed an approach with Blockchain based secure communication scheme and an interplanetary file system over the 6G network. They presented its challenges and future directions.

Due to the ease of customization and synchronization to several computing applications, providing secure communication in drones is a challenging task. The communication nodes are prone to coagulation attacks, resulting from the clotting properties of fluids. Srinivasan et al. [121] discussed coagulation attacks and their impact on various underwater exploration tasks. The authors highlighted the importance of the field of UWC for the transmission of data through optical, acoustic and electromagnetic waves. The common underwater communication modes, acoustic, optical and radiofrequency suffer from issues. Acoustic allows long-range communication (in Kms), however, provides low data throughput, making the overall process time-consuming. Optical needs precise alignment between sender and receiver for transmission. Radiofrequency offers high speed but a short range (between a few centimetres to a few meters). There is a need for broadband access to facilitate long-range underwater transmission. The article "Wireless Underwater Broadband and Long Range Communications using underwater Drones as Data Mules' [122] proposed the use of small-sized underwater drones, named data mules, and the expansion of their range using RF transmissions. They used communication protocols with connectivity constraints in Delay Tolerant Networks, demonstrated raised throughput, in the experimental setup, and compared with existing commercial acoustic modems. They introduced changes in source code for bidirectional synchronization of files and presented extrapolation of results for higher-distance underwater communication.

One of the significant applications of UWC techniques is an early warning and disaster prevention during tsunamis. Srinivasan et al. [121] presented an overview to UWC, along with challenges, research directions and recommendations using 5G networks. To meet the Quality of Service requirements through high data rates and low latency, 5G networks are useful, especially for large deployments of self-driving UAVs/UUDs. 5G coupled with fog and MEC provides high data rate results for complex autonomous vehicles.

Large smart city scenarios with an increasing number of connected devices, suffer from irregular data and service requests, both in dense and less sparse locations. There is a need for fixed base stations that turn out to be expensive. Blockchain-integrated UUDs enabled as on-demand access points serve dynamic user requests reliably and securely and offer high data availability due to fog and cloud computing devices. As against a UUD supported by traditional cellular networks, fog and cloud-supported UUDs provide improved data delivery and message-exchange success rates. Aloqaily [123] discussed emerging technologies such as federated learning, challenges and future research opportunities.

Traditionally, the communication between the ground control station and UUD is through communication protocols, such as MAVLink, UranusLink, UAVCAN, which offer considerable security. However, these are prone to MITM, DoS attacks, packet data injection and eavesdropping. This compromises the security of messages that bear considerable information about UUD, in the form of control commands sent to and from UUDs and ground stations. Consequences range from crash landing to stealing of sensitive information (in case of military operation), and false injection of reports (in case of an investigation or search and rescue operation). Advocating a need for a secure communication protocol, with requisite security standard sets, Khan et al. [124] presented a survey covering general architecture, attacks and proposed a novel communication protocol for UUD security issues. Table 7 presents some recent research in the context of blockchain integration and its limitations thereof.

Table 7: Recent researches on blockchain integration and limitations

| Authors | Year | Approach & Key strength | Limitation |
|---|---|---|---|
| Gupta et al. [112] | 2021 | Blockch-based communication scheme over 6G is proposed, highlighting its advantages around security, privacy and low storage cost. | However, comparison with other existing communication schemes in 5G, 6G solutions is not explored |
| Srinivasan et al. [121] | 2019 | UAVs are introduced as significant vehicles of maritime research. The importance of UWC is highlighted, presenting challenges, emerging technologies in UWC, and future directions using 5G communication techniques | Integration of blockchain in UWC and insights thereof are not discussed. |
| Aloqaily et al. [122] | 2021 | a 5G Network environment supported by blockchain-enabled UAVs, to manage dynamic user needs, is discussed. They highlighted their approach with secure and reliable routing, authentication and higher availability through fog and cloud | Detailing federated learning in the context of blockchain is not explored. |
| | | elements integrated. A comparison of success rates with cellular networks was detailed. | |
| Khan et al. [123] | 2020 | Security attacks on UAVs are discussed, and the need for secure communication protocol is highlighted. | A comprehensive review of a variety of UAV attacks and insights into existent communication protocols and their vulnerabilities in protecting sensitive data is discussed, however, no practical implementation or blockchain integration is discussed. |

Other core directions, ideas, and paradigms that support blockchain and future-generation network integration for UUDs are briefly discussed as follows[112][121][122][123]:

- Recent years have seen the development of innovative approaches to underwater communication that make use of the many advantages offered by 6G. Both the Basic Bio-Inspired Camouflage Communication Frame (BBICF) and the Ultra Violet Light Cannon (UVLC) will be of tremendous use in future applications. These innovative strategies were developed to maximise the advantages that 6G technology has to offer. There is a sizeable amount of the Earth's surface that is covered with water, and this quantity represents a sizeable proportion of the overall land area of the planet. The oceans are to responsible for a disproportionate amount of the general rise in temperature that has been seen on Earth. This is because the oceans retain vast amounts of heat, which has caused them to contribute to the warming trend. Because of the importance of the ocean to human life, there is an immediate and compelling need to conduct research and exploration of the ocean. For example, the wealth of information obtained by underwater surveys has made the seafloor the focus of a significant amount of investigation and discovery over the last 10 years. This is mostly because researchers have been able to access more of the seafloor. This has resulted in the finding of a considerable number of previously unknown creatures. Interaction with other people and the dissemination of knowledge are both made simpler by high-tech networks that are conceived with the future in mind (such as 6G or above).
- Many different kinds of intelligent industries, like as healthcare, distant education, IoT, autonomous cars, and ultra-smart homes and cities, stand to gain from the introduction of 5G. Holographic communication, five-sense communication, and wireless brain-computer linkages are some of the potential forms of touch that researchers are looking at for the not-too-distant future. The elimination of the data gap that has previously existed between the real world and digital networks of communication is one of the primary contributions that the Internet of Things makes to the development of these cutting-edge technological applications [147][148]. This data gap has been a major obstacle in the past. For example, the term "holographic-type communications" (HTC) refers to a new type of communication in which holographic applications send holographic data to a remote location while real holograms are created at the receiving end of the transmission. HTC opens the way for the manufacturing of holograms on-site by simplifying the connections necessary to communicate various service data flows from a large number of view cameras. These connections are required to communicate numerous service data flows. This is one of how HTC is paving the way for the development of holograms. The hologram has the potential to imitate the behaviours and feelings of a real person if it is shown on a screen that is capable of holographic projection. Sadly, this indicates that HTC is not utilised in 5G by a large number of people. These technologies may be easily included into 6G networks using a variety of radio access methods. In contrast, 5G systems are made up of many networks that connect at different depths and in different ways, including the traditional terrestrial, aerial, satellite, and subsea layers. 5G networks is a common shorthand for these kinds of triads. It is also expected that 6G networks would widely disperse intelligent computing and communication across these heterogeneous access networks, especially at the network's periphery. Here, blockchain technology makes it possible for networks to be operated in a decentralised, dynamic, and low-cost manner. Intelligent computing and communication devices have a better chance of communicating with one another if they use blockchain's centralised authentication technique. When data is spread out among several nodes, there is no longer a single vulnerable node that might cause problems. This is because the data that is stored using blockchain technology is encrypted, it is difficult for unauthorised parties to see the data or make any modifications to it.
- Geng et al. [148] recommended that underwater acoustic networks use a protocol called deep-learning-based media access control (DL-MAC) in order to improve the general quality of their communications. In order to improve the effectiveness of these networks, this measure is essential . The DL-MAC agent was developed to maximise the utilisation of available bandwidth despite the considerable delays in propagation that are characteristic of underwater communications. This was done so that the available bandwidth could be maximised despite the obstructions. Both synchronous and asynchronous methods of data transfer may achieve this goal. Both the synchronous and asynchronous protocols have been proved to be effective via a mix of theoretical research and computational simulations. Both synchronous and asynchronous machine learning agents use deep Q-learning to understand which behaviours are most productive. This is done so that the best possible learning environment may be found. Blockchain technology due to its distributed and decentralized approach can ensure high security standards for this mechanism. It can keep track of performance statistics more accurately and provide high security of data at its transmission, propagation and storage stages.

### 7.4. Blockchain, Fog, Edge and Serverless Computing for UUDs

UUD with MEC connects to IoUT and executes several crucial underwater surveillance and monitoring tasks. The communication of UUD from remote areas with IoUT is prone to cyberattacks. Integration with Blockchain technology provides support in terms of security and integrity. Islam and Shin [125] presented a Blockchain based secure scheme for UUD, integrated with MEC, and described a use case of underwater monitoring. In the near future, drones shall accomplish several complex drone missions due to their ability to work in collaboration. Capability of drones to work in collaborative drone missions depends greatly on reduction in communication complexity and in ability to be controlled in a decentralized manner. Blockchain provides decentralized data, and facilitates public access to all neighboring drones to log information related to the mission, such as, location, state of delivery, time, resources etc. Alsamhi et al. [126] introduced decentralized multi-drones to execute collaborative drone missions. They proposed improvement in Blockchain through consensus algorithms to improve network partitioning and scalability. Uddin et al. [29] proposed a multilevel sensor monitoring architecture, with fog and edge elements to store and process the data captured from IoUT securely through Blockchain, and ensuring immutability and irreversibility of data. The authors demonstrated the architecture presenting secure routing through its hierarchical topology and discussed findings, indicating its efficient and secure functioning.

#### 7.4.1. Integration of Blockchain with Fog and Edge elements and UUD Edge networks

Due to growing latency and cost issues of large cloud servers, the use of edge cloud has become prominent in UUD Networks. Close proximity to connected devices ensures high bandwidth, low turbulence and owner's control in terms of enforcing privacy policies on edge servers before releasing data to the cloud. MEC is one of the key technologies towards 5G. An important use case of UUD networks implemented with MEC, is in disasters, where stationary ground infrastructure may not be available. It offers availability and robustness. Blockchain integration facilitates maintaining high trust between participant UUDs in UUD networks. Using UUD caching, Blockchain can increase reliability.

The application services on edge servers should be migrated closer to the edge nodes, to ensure higher quality services. Zang et al. [127] targeted edge service migration in the MEC environment to improve user service quality. They presented a Blockchain based edge service migration framework, named Falcon. They demonstrated, use of mobile agents and service migration algorithms to maximize benefits of migration and yield high service quality.

Sharma et al. [128] presented an ultra-reliable caching scheme in Edge-UUD networks with Blockchain for increasing reliability of MEC communication. They presented a neural network model to generate master Blockchain to predict the optimal path for UUD. As an extension of the same, they presented a neural network approach to compute the reliability parameter of UUD Network. Using smart contracts and transaction models, Blockchain facilitates sharing of content and data between UUD, edge servers and caching servers. Deriving page rank from survivability of page requests from UUDs, Blockchain supports caching. Use of Blockchain delivers ultra-reliable MEC communications, with increased probability of connectivity, survivability, reliability and reduced energy consumption .

Álvares et al. [129] presented several challenges in UUDs concerning lower and upper bounds on memory for Blockchain-enabled communication, survivability issues in case of UUDs with voluminous computations, and other challenges including failure analysis, server dominance, content sharing and identification. Some of the solutions are partitioning and offloading in smart mobile devices, which when adopted add to suitability of Blockchain based UUD networks for MEC applications.

Security and scalability are crucial requirements for Industrial IOT and critical infrastructures of Industry 4.0. Large volumes of data collected and analyzed place huge computing demands that edge computing can handle.

Integration of Blockchain in IIOT, serves to handle security issues. Wu et al. [130] presented a survey on confluence of Blockchain and edge computing paradigms. They introduced a layered architecture for IIOT critical infrastructures, for effective handling of security, privacy and scalability issues, and highlighted open research challenges.

In keeping with high computational cost and transmitting huge volumes of sensor data securely and privately across hierarchical sensor networks remains a challenge, and fog and cloud elements integrated with Blockchain offer valuable solutions. Uddin et al. [29] proposed a Blockchain integrated multilevel sensor monitoring architecture, with fog and cloud elements to process data captured by sensors at varying depths, to transfer to base stations at the surface, securely. Effective performance results through secure routing as part of the architecture were demonstrated.

### 7.4.2. Confluence of Blockchain and Serverless Computing

Yuan [131] stated, Serverless computing eliminates the need to provision servers and manage hardware resources. With Serverless, the code is easy to deploy for developers. FAAS is becoming popular with AWS Lambda, Azure functions, and Amazon API gateway, available to code, without need to bother about operating system, hardware and virtual machines. The developer is free to code, and it provisions and scales as per demand. The compute request of each of the applications is serviced individually and precisely, allowing for efficient scaling, as per the size of the workload. This brings considerable savings in terms of cost, as the payment is as per use. Taking a case in focus, the Serverless computing service, such as, AWS Lambda, runs the code, and automatically handles compute resources for the developer, in event of change of data, shift of state, or an action by user, as in case of a data processing application.

FAAS requires containers such as Docker to deploy the function code. This has issues such as slow processing, due to cold start problems, and poses overhead due to the need for an entire OS and stack, only for the purpose of running a single function. This further also adds on to the management of server resources. First generation Blockchains, being decentralized ledgers, allow untrusted parties to reach consensus, using PoW consensus algorithm. One of the best known cryptocurrencies is bitcoin. The second generation Blockchains (such as Ethereum) work more like general computing platforms, and allow untrusted parties to run the same code as functions/smart contracts, with the consensus being reached by a set of complex consensus protocols. There is no concept of a server, and the code is run by the entire Blockchain network, which charges a 'gas fee' in cryptocurrency for the computation. The results are placed in the next Blockchain. Based on the ops code in the compiled byte code and the storage space consumed, the 'gas fee' is computed. This is highly precise than any of the traditional cloud service providers available in current times. The performance measure of Blockchain is, how quickly it can reach agreements in consensus protocols [131].

Sharma [132] stated that, while Blockchain and serverless technologies are distinctly different from each other, the convergence has manifold benefits, in terms of building resistant apps, maintaining trust and saving costs. Blockchain is stateful, making it difficult to scale (the smart contracts suffer with scalability issues, to even the extent of 1000 transactions per second), whereas, serverless is stateless. While the Blockchains are persistent, the serverless computation is short lived, scales well, but relies on trust. Trust is inbuilt into the components of Blockchain architecture, through consensus mechanisms, to ensure validation of transactions and safety. Blockchain provides a public and an independently verifiable way to maintain transactional states for stateless computation. Blockchain can share the trust with outsiders as well]. Table 8 presents some of the recent research in blockchain integration and their limitations thereof.

Table 8: Recent research in blockchain integration and limitations.

| Authors | Year | Approach & Key strength | Limitation |
|---|---|---|---|
| Zang et al. [127] | 2019 | Blockchain based secure edge service migration framework 'Falcon' is proposed, with better service quality and flexibility. Immutable alliance chain among multiple edge clouds is discussed, to ensure reliable and safe edge service. Experiments demonstrating low energy consumption are detailed. | Comparison with other contemporary service migration frameworks is not detailed. |
| Sharma et al. [128] | 2019 | Factors related to ultra reliable communication in MEC, when using drones as on demand nodes, are covered. Focus is on reliability through a novel neural blockchain based drone caching approach presented. | While flat architecture focusing on reliability factor is emphasized, the multi drone scenarios and impact of other factors such as service quality is not explored. |
| Alvares et al. [129] | 2021 | Resource constraints of IoV networks, when integrated with Blockchain and the challenges around computational and energy requirements thereof have been discussed | However, solutions in term of light weight blockchain integrations for smart city implementations have not been explored. |
| Wu et al. [130] | 2021 | Convergence of edge and blockchain paradigms to enable secure data handling in critical infrastructures of Industry 4.0 is discussed | However, only introducing the two paradigms and possible convergence is deliberated. Detailed use cases and implications are not discussed. |

## 6. OPEN RESEARCH CHALLENGES AND FUTURE DIRECTIONS

Overall, the extant literature suggests that there are critical challenges for UUD implementation for commercial and civil use. For example, the salient knowledge gaps and opportunities are related to preventing collisions in sea activities, managing UUD swarms in oceans, cyber-physical attacks on UUD and pathway plans could be potentially addressed by Blockchain applied in isolation or combined with other disruptive technologies. Moreover, the existing Blockchain is not totally efficient to process IoUT big data [29]. Therefore, due to the early stage of the research in discussing Blockchain in UUD applications, there has been little empirical and potential of generalization of the results.

In a nutshell, some of the current challenges for realistic use of Blockchain that reduce the performance and storage capacities include high computing costs and delays [30]. Moreover, there has been little research on the practical design of control techniques that consider performance, formation control, and underwater communication capabilities [25] [26]. In this case, Blockchain protocols can make significant scientific contributions. It is also essential to evaluate AUV coordination and formation from an interdisciplinary perspective to assist researchers in avoiding unrealistic research [25].

Because of the restricted battery life of commercially available UAVs, the flight duration remains an issue. Consequently, In this situation, sophisticated optimization algorithms for UAV operations and routes are required to reduce unnecessary trajectories and battery consumption. Similarly, adopting Blockchain for UUD will necessitate additional processing power, increasing energy consumption. Therefore, more research is needed to create new technologies and practical answers to increase the flying duration of UUDs adopting Blockchain in maritime operations [23][135][136][137][138][139]. Additionally, empirical research demonstrating applications of UUD for acoustic monitoring using robotics is needed [27]. Also, studies designing robust recovery mechanisms when coordinated groups of UUD failures are missing [25].

The state-of-the-art of literature indicates that more research is also needed to turn the private Blockchain networks more secure and unchangeable because they are more exposed to cyber-attacks than public Blockchains networks [23] [140][141][142]. In addition, the environmental integrity of water systems is generally assessed using single-point measurements that ignore the time and geographic variability of the water systems. This implies that more dynamic and robust monitoring technologies and methods are required [28], and research in this direction is needed. In short, our paper analyzes the key limitations for effective Blockchain integration in UUD applications, demonstrates the main benefits, and adds new information to the existing literature on these growing subjects. In addition, the article provides suggestions for advancements and future researches [143][144][145].

## 7. CONCLUSIONS

The study reviewed state-of-art and knowledge gaps in integration of Blockchain in real time under water drone applications, underscoring importance of Blockchain as a solution to several data security and privacy concerns. Background and technical issues were discussed. Effort in direction to add new knowledge to this evolving field has been made. Research work is needed to convert private Blockchain networks to more robust and unchangeable, to protect against cyberattacks. Underwater terrains present challenges demanding unique implementations and solutions. There are limitations in effective integration of Blockchain in realistic implementations of underwater drone applications opening up avenues for future research. Integration of Blockchain with UUD configurations, warrants studies on computational testing, summation models, new tools, and frameworks. Advances in computing power, improved battery systems, use of advanced sensors, and integration of smart algorithms, improved software models and collision avoidance frameworks are various aspects of ongoing developments in UUDs. Developments in on-board intelligence, abilities of sensing, navigation and autonomy are expected, to unlock many secrets of oceans, contribute to high quality research, tackle sophisticated tasks, and serve variety of purposes including waging of wars.

## Appendix A: List of Abbreviations

| Acronym | Abbreviation |
|---|---|
| AAV | Autonomous Aerial Vehicles |
| AI | Artificial Intelligence |
| ANN | Artificial Neural Network |
| AR | Augmented Reality |
| AUVs | Autonomous Underwater Vehicles |
| BC-VNDN | Blockchain based vehicular Named Data Network |
| CNN | Convolutional Neural Network |
| CPS | Cyber-Physical Systems |
| DAG | Distributed Acyclic Graph |
| DO | Dissolved Oxygen |
| DoS | Denial of Service |
| DANNC | Distributed Adaptive Neural Network Control |
| FAAS | Function-As-A-Service |
| GIMP | GNU Image Manipulation Program |
| GPS | Global Positioning System |
| HFWUG | High-efficiency Flying Wing Underwater Glider |
| ICT | Information Communication Technologies |
| IoD | Internet of Drones |
| IoT | Internet of Things |
| IIoT | Industrial Internet of Things |
| IoMT | Internet of Medical Things |
| IoUT | Internet of Underwater Things |
| IPFS | The InterPlanetrary File System |
| KNN | K Nearest Neighbour |
| LED | Light Emitting Diode |
| MEC | Mobile Edge Computing |
| MITM | Man-In-The-Middle |
| ML | Machine Learning |
| MVS | Multiple view Stereo |
| NACA | National Advisory Committee for Aeronautics |
| OBIA | Object Based Image Analysis |
| ORP | Oxidation Reduction Potential |
| PBFT | Practically Byzantine Fault Tolerance |
| PoA | Proof of Authority |
| PoET | Proof of Elapsed Time |
| POGO | Partnership for Observation of the Global Oceans |
| PoP | Proof of PUF-Enabled Authentication |
| PoS | Proof of Stake |
| PoV | Proof of Vote |
| PoW | Proof of Work |
| ROD | Remotely Operated Device |
| RODr | Remotely Operated Drones |
| ROS | Robot Operating System |
| ROVs | Remotely Operated Vehicles |
| SFM | Structure from Motion |
| SONAR | Sound NAvigation and Ranging |
| SVM | Support Vector Machine |
| TWSNs | Territorial Wireless Sensor Networks |
| UAV | Unmanned Aerial Vehicle |
| UGV | Unmanned Ground Vehicle |
| USV | Unmanned Surface Vehicle |
| UUD | Unmanned Underwater Drone |
| UUV | Unmanned Underwater Vehicles |
| UWC | Underwater Wireless Communication |
| UWSNs | Underwater Wireless Sensor Networks |
| VR | Virtual Reality |
| WHO | World Health Organization |